\documentclass[10pt,a4paper]{article}
\usepackage[latin1]{inputenc}
\usepackage[english]{babel}
\usepackage{amsfonts}
\usepackage{amssymb}
\usepackage{amsmath}
\usepackage{latexsym}
\usepackage[T1]{fontenc}
\usepackage{subfig}
\usepackage{tikz}
\newtheorem{theoreme}{Theorem}[section]
\newtheorem{lemme}[theoreme]{Lemma}

\newtheorem{proposition}[theoreme]{Proposition}
\newtheorem{remarque}[theoreme]{Remark}

\newenvironment{preuve}{\emph{Proof} : }{\begin{flushright}$\Box$\end{flushright}}
\newcommand{\F}{\mathbb{F}}

\title{On the third weight of generalized Reed-Muller codes}
\author{Elodie Leducq}
\date{} 
\begin{document}
\maketitle

\begin{abstract}In this paper, we study the third weight of generalized Reed-Muller codes. Using results from \cite{erickson1974counting}, we prove under some restrictive condition that the third weight of generalized Reed-Muller codes depends on the third weight of generalized Reed-Muller codes of small order with two variables. In some cases, we are able to determine the third weight and the third weight codewords of generalized Reed-Muller codes.\end{abstract}

\section{Introduction}

In this paper, we study the third weight of generalized Reed-Muller codes.\\

We first introduce some notations : \\

Let $p$ be a prime number, $e$ a positive integer, $q=p^e$ and $\F_q$ a finite field with $q$ elements.\\

If $m$ is a positive integer, we denote by $B_m^q$ the $\F_q$-algebra of the functions from $\F_q^m$ to $\F_q$ and by $\F_q[X_1,\ldots,X_m]$ the $\F_q$-algebra of polynomials in $m$ variables with coefficients in $\F_q$. 

We consider the morphism of $\F_q$-algebras $\varphi: \F_q[X_1,\ldots,X_m]\rightarrow B_m^q$ which associates to $P\in\F_q[X_1,\ldots,X_m]$ the function $f\in B_m^q$ such that $$\textrm{$\forall x=(x_1,\ldots,x_m)\in\F_q^m$, $f(x)=P(x_1,\ldots,x_m)$.}$$ The morphism $\varphi$ is onto and its kernel is the ideal generated by the polynomials $X_1^q-X_1,\ldots,X_m^q-X_m$. So, for each $f\in B_m^q$, there exists a unique polynomial $P\in\F_q[X_1,\ldots,X_m]$ such that the degree of $P$ in each variable is at most $q-1$ and $\varphi(P)=f$. We say that $P$ is the reduced form of $f$ and we define the degree $\deg(f)$ of $f$ as the degree of its reduced form.
 The support of $f$ is the set $\{x\in\F_q^m:f(x)\neq0\}$ and we denote by $|f|$ the cardinal of its support (by identifying canonically $B_m^q$ and $\F_q^{q^m}$, $|f|$ is actually the Hamming weight of $f$).
\\

For $0\leq r\leq m(q-1)$, the $r$th order generalized Reed-Muller code of length $q^m$ is $$R_q(r,m):=\{f\in B_m^q :\deg(f)\leq r\}.$$

For $1\leq r\leq m(q-1)-2$, the automorphism group of generalized Reed-Muller codes $R_q(r,m)$ is the affine group of $\F_q^m$ (see \cite{charpin_auto}).
\\

 For more results on generalized Reed-Muller codes, we refer to \cite{delsarte_poids_min}.
\\

In the following of the article, we write $r=a(q-1)+b$, $0\leq a\leq m-1$, $1\leq b\leq q-1$. \\

In \cite{MR0275989}, interpreting generalized Reed-Muller codes in terms of BCH codes, it is proved that the minimal weight of the generalized Reed-Muller code $R_q(r,m)$ is $(q-b)q^{m-a-1}$. The minimum weight codewords of generalized Reed-Muller codes are described in \cite{delsarte_poids_min} (see also \cite{Leducq2012581}).

In his Ph.D thesis \cite{erickson1974counting}, Erickson proves that if we know the second weight of $R_q(b,2)$, then we know the second weight for all generalized Reed-Muller codes. From a conjecture on blocking sets, Erickson conjectures that the second weight of $R_q(b,2)$ is $(q-b)q+b-1$. Bruen proves the conjecture on blocking set in \cite{MR2766082}. Geil also proves this result in \cite{MR2411119} using Groebner basis. An altenative approach can be found in \cite{MR2592428} where the second weight of most $R_q(r,m)$ is established without using Erickson's results. Second weight codewords have been studied in \cite{MR1384161}, \cite{MR2332391} and finally completely described in \cite{raey}.

For $q=2$, small weights and small weight codewords are described in \cite{MR0401324}, the third weight for $r>(m-1)(q-1)+1$ is given in \cite{MR2411119}, we can find results on small weight codewords in \cite{DBLP:journals/corr/abs-1203-4592}. In the following, we consider only $q\geq3$ and $r\leq m(q-1)+1$.\\

We first give some tools that we will use through all the paper. Then we give an upper bound on the third weight of generalized Reed-Muller codes. In Section 4 is the main result of this article : we describe the third weight of generalized Reed-Muller codes with some restrictive conditions. In section 5, we study more particularly the case of two variables which is quite essential in the determination of the third weight. In Section 6, we described the codewords reaching the third weight. In section 7, we summarize the results obtain in this article. This article ends with an Appendix which gives more precisions on the results in Section 3.  

\section{Preliminaries}

\subsection{Notation and preliminary remark}

Let $f\in B_m^q$, $\lambda\in\F_q$. We define $f_{\lambda}\in B_m^q$ by $$\forall x=(x_2,\ldots,x_m)\in\F_q^m, \quad f_{\lambda}(x)=f(\lambda,x_2\ldots,x_m).$$

Let $0\leq r\leq (m-1)(q-1)$ and $f\in R_q(r,m)$. We denote by $S$ the support of $f$. Consider $H$ an affine hyperplane in $\F_q^m$, by an affine transformation, we can assume $x_1=0$ is an equation of $H$. Then $S\cap H$ is the support of $f_0\in R_q(r,m-1)$ or the support of $(1-x_1^{q-1})f\in R_q(r+(q-1),m)$.

\subsection{Useful lemmas}

\begin{lemme}\label{inter}Let $q\geq 3$, $m\geq 3$ and $S$ a set of points of $\F_q^m$ such that $\#S=uq^n<q^m$, $u\not\equiv0\mod q$. If for all hyperplane $H$ $\#(S\cap H)=0$, $\#(S\cap H)=wq^{n-1}$, $\#(S\cap H)=vq^{n-1}$ or $\#(S\cap H)\geq uq^{n-1}$, with $w<v<u$, then there exists $H$ an affine hyperplane such that $\#(S\cap H)=0$, $\#(S\cap H)=wq^{n-1}$ or $\#(S\cap H)=vq^{n-1}$.\end{lemme}

\begin{preuve}Assume for all $H$ hyperplane, $\#(S\cap H)\geq uq^{n-1}$. Consider an affine hyperplane $H$; then for all $H'$ hyperplane parallel to $H$, $\#(S\cap H')\geq u.q^{n-1}.$ 
Since $u.q^{n}=\#S=\displaystyle\sum_{H'//H}\#(S\cap H')$, we get that for all  $H$ hyperplane, $\#(S\cap H)=u.q^{n-1}$.
\\Now  consider $A$ an affine subspace of codimension 2 and the $(q+1)$ hyperplanes through $A$. These hyperplanes intersect only in $A$ and their union is equal to $\F_q^m$. So $$uq^{n}=\#S=(q+1)u.q^{n-1}-q\#(S\cap A).$$
Finally we get a contradiction if $n=1$ since $u\not\equiv 0\mod q$. Otherwise, $\#(S\cap A)=u.q^{n-2}$. Iterating this argument, we get that for all $A$ affine subspace of codimension $k\leq n$, $\#(S\cap A)=u.q^{n-k}$. 
\\Let $A$ be an affine subspace of codimension $n+1$ and  $A'$ an affine subspace of codimension $n-1$ containing $A$. We consider the $(q+1)$ affine subspaces of codimension $n$ containing $A$ and  included in $A'$, then $$u.q=\#(S\cap A')=(q+1)u-q\#(S\cap A)$$ which is absurd since $\#(S\cap A)$ is an integer and $u\not\equiv 0\mod q$. So there exists $H_0$ an hyperplane such that $\#(S\cap H_0)=vq^{n-1}$, $\#(S\cap H_0)=wq^{n-1}$ or $S$ does not meet $H_0$. \end{preuve}

The following lemma is proved in \cite{delsarte_poids_min}. 

\begin{lemme}\label{DGMW1}Let $m\geq1$, $q\geq2$, $f\in B_m^q$ and $w\in\F_q$. If for all $(x_2,\ldots,x_m)$ in $\F_q^{m-1}$, $f(w,x_2,\ldots,x_m)=0$ then for all $(x_1,\ldots,x_m)\in\F_q^m$, $$f(x_1,\ldots,x_m)=(x_1-w)g(x_1,\ldots,x_m)$$ with $\deg_{x_1}(g)\leq\deg_{x_1}(f)-1$ and $\deg(g)\leq\deg(f)-1$.\end{lemme}

The following lemmas are proved in \cite{erickson1974counting}

\begin{lemme}\label{2.7}Let $m\geq 2$, $q\geq3$, $0\leq r\leq m(q-1)$. If $f\in R_q(r,m)$, $f\neq0$ and there exists $y\in R_q(1,m)$ and $(\lambda_i)_{1\leq i\leq n}$ $n$ elements in $\F_q$ such that the hyperplanes of equation $y=\lambda_i$ do not meet the support of $f$, then $$|f|\geq(q-b)q^{m-a-1}+\left\{\begin{array}{ll}n(b-n)q^{m-a-2}&\textrm{if $n<b$}\\(n-b)(q-1-n)q^{m-a-1}&\textrm{if $n\geq b$}\end{array}\right.$$ where $r=a(q-1)+b$, $1\leq b\leq q-1$.\end{lemme}

\begin{lemme}\label{3.9}Let $m\geq2$, $q\geq3$, $1\leq b\leq q-1$. Assume $f\in R_q(b,m)$ is such that $f$ depends only on $x_1$ and $g\in R_q(b-k,m)$, $1\leq k\leq b$. Then either $f+g$ depends only on $x_1$ or $|f+g|\geq (q-b+k)q^{m-1}$.\end{lemme}

\begin{lemme}\label{3.5}Let $m\geq2$, $q\geq3$, $1\leq a\leq m-1$, $1\leq b\leq q-2$. Assume $f\in R_q(a(q-1)+b,m)$ is such that $\forall x=(x_1,\ldots,x_m)\in\F_q^m$, $$f(x)=(1-x_1^{q-1})\widetilde{f}(x_2,\ldots,x_m)$$ and $g\in R_q(a(q-1)+b-k)$, $1\leq k\leq q-1$, is such that $(1-x_1^{q-1})$ does not divide $g$. Then, either $|f+g|\geq (q-b+k)q^{m-a-1}$ or $k=1$.\end{lemme}


\begin{lemme}\label{3.7}Let $m\geq2$, $q\geq3$, $1\leq a\leq m-2$, $1\leq b\leq q-2$ and $f\in R_q(a(q-1)+b,m)$. We set an order on the elements of $\F_q$ such that $|f_{\lambda_1}|\leq\ldots\leq|f_{\lambda_q}|$.

If $f$ has no linear factor and there exists $k\geq1$ such that $(1-x_2^{q-1})$ divides $f_{\lambda_i}$ for $i\leq k$ but $(1-x_2^{q-1})$ does not divide $f_{\lambda_{k+1}}$ then, $$|f|\geq (q-b)q^{m-a-1}+k(q-k)q^{m-a-2}.$$\end{lemme}

\begin{lemme}\label{2.14}Let $m\geq2$, $q\geq3$, $1\leq a\leq m$ and $f\in R_q(a(q-1),m)$ such that $|f|=q^{m-a}$ and $g\in R_q(a(q-1)-k,m)$, $1\leq k\leq q-1$, such that $g\neq0$. Then, either $|f+g|=kq^{m-a}$ or $|f+g|\geq(k+1)q^{m-a}$.\end{lemme}

\begin{lemme}\label{2.15.1}Let $m\geq2$, $q\geq3$, $1\leq a\leq m-1$ and $f\in R_q(a(q-1),m)$. If for some $u$, $v\in\F_q$, $|f_u|=|f_v|=q^{m-a-1}$, then there exists $T$ an affine transformation fixing $x_1$ such that $$(f\circ T)_u=(f\circ T)_v.$$\end{lemme}

\section{An upper bound on the third weight}\label{upper}

\begin{theoreme}\label{3hyp}Let $q\geq 3$, $m\geq2$, $0\leq a\leq m-1$, $1\leq b\leq q-1$, then if $W_3$ is the third weight of $R_q(a(q-1)+b,m)$, we have
\begin{itemize}
\item If $b=1$ then,
\begin{itemize}\item For $q=3$, $m\geq3$, $1\leq a\leq m-2$, $$W_3\leq3^{m-a}.$$
\item For $q=4$, $m\geq3$ and $1\leq a\leq m-2$, $$W_3\leq18.4^{m-a-2}.$$
\item For $q=3$ and $a=m-1$ or $q=4$ and $a=m-1$, $$W_3\leq2(q-1).$$
\item For $q\geq5$ and $1\leq a\leq m-1$, 
$$W_3\leq2(q-2)q^{m-a-1}.$$
\end{itemize}
\item If $2\leq b\leq q-1$
\begin{itemize}
\item For $q\geq5$, $0\leq a\leq m-2$ and $4\leq b\leq \lfloor\frac{q}{2}+2\rfloor$,
$$W_3\leq(q-2)(q-b+2)q^{m-a-2}.$$
\item For $q\geq7$, $0\leq a\leq m-2$ and $\lceil\frac{q}{2}+2\rceil\leq b\leq q-1$ or $q\geq 4$, $0\leq a\leq m-2$ and $b=2$ or $q\geq 4$, $a=m-2$ and $b=3$ or $q=3$, $a\in\{0,m-2\}$ and $b=2$ 
$$W_3\leq(q-b+1)q^{m-a-1}.$$
\item For $q\geq4$, $m\geq3$, $0\leq a\leq m-3$ and $b= 3$,
$$W_3\leq(q-1)^3q^{m-a-3}.$$
\item For $q=3$, $m\geq4$, $1\leq a\leq m-3$ and $b=2$, $$W_3\leq16.3^{m-a-3}.$$
\end{itemize}
\end{itemize}
\end{theoreme}

\begin{preuve}\begin{itemize}
\item If $b=1$ then,
\begin{itemize}\item For $q=3$, $m\geq3$, $1\leq a\leq m-2$, define for $x=(x_1,\ldots,x_m)\in\F_q^m$, $$f(x)=\prod_{i=1}^a(1-x_i^{2}).$$ Then, $f\in R_3(2a+1,m)$ and $|f|=3^{m-a}>8.3^{m-a-2}$.
\item For $q=4$, $m\geq3$ and $1\leq a\leq m-2$, define for $x=(x_1,\ldots,x_m)\in\F_q^m$, $$f(x_1,\ldots,x_m)=\prod_{i=1}^{a-1}(1-x_i^3)(x_a-u)(x_a-v)(x_{a+1}-w)(x_{a+2}-z)$$ with $u$, $v$, $w$, $z\in\F_q$ and $u\neq v$. Then, $f\in R_4(3a+1,m)$ and $|f|=18.4^{m-a-2}>4^{m-a}$
\item For $q=3$ and $a=m-1$ or $q=4$ and $a=m-1$, define for $x=(x_1,\ldots,x_m)\in\F_q^m$, $$f(x)=\prod_{i=1}^{m-2}(1-x_i^{q-1})\prod_{j=1}^{q-2}(x_{m-1}-b_j)(x_{m}-c)$$ with $b_j\in\F_q$, $b_j\neq b_k$ for $j\neq k$ and $c\in\F_q$. Then, $f\in R_q((m-1)(q-1)+1,m)$ and $|f|=2(q-1)>q$
\item For $q\geq5$ and $1\leq a\leq m-1$, define for $x=(x_1,\ldots,x_m)\in\F_q^m$,
$$f(x)=\prod_{i=1}^{a-1}(1-x_i^{q-1})\prod_{j=1}^{q-2}(x_a-b_j)(x_{a+1}-u)(x_{a+1}-v)$$ with $b_j\in\F_q$, $b_j\neq b_k$ for $j\neq k$ and $u$, $v\in\F_q$, $u\neq v$. Then, $f\in R_q(a(q-1)+1,m)$ and $|f|=2(q-2)q^{m-a-1}>q^{m-a}$
\end{itemize}
\item If $2\leq b\leq q-1$
\begin{itemize}
\item For $q\geq5$, $0\leq a\leq m-2$ and $4\leq b\leq \lfloor\frac{q}{2}+2\rfloor$, define for $x=(x_1,\ldots,x_m)\in\F_q^m$,
$$f(x)=\prod_{i=1}^a(1-x_i^{q-1})\prod_{j=1}^{b-2}(x_{a+1}-b_j)(x_{a+2}-u)(x_{a+2}-v)$$ with $b_j\in\F_q$, $b_j\neq b_k$ for $j\neq k$ and $u$, $v\in\F_q$, $u\neq v$. Then, $f\in R_q(a(q-1)+b,m)$ and $|f|=(q-2)(q-b+2)q^{m-a-2}>(q-b+1)(q-1)q^{m-a-2}$
\item For $q\geq7$, $0\leq a\leq m-2$ and $\lceil\frac{q}{2}+2\rceil\leq b\leq q-1$ or $q\geq 4$, $0\leq a\leq m-2$ and $b=2$ or $q\geq 4$, $a=m-2$ and $b=3$ or $q=3$, $a\in\{0,m-2\}$ and $b=2$, define for $x=(x_1,\ldots,x_m)\in\F_q^m$, 
$$f(x)=\prod_{i=1}^a(1-x_i^{q-1})\prod_{j=1}^{b-1}(x_{a+1}-b_j)$$ with $b_j\in\F_q$, $b_j\neq b_k$ for $j\neq k$. Then, $f\in R_q(a(q-1)+b,m)$ and $|f|=(q-b+1)q^{m-a-1}>(q-b+1)(q-1)q^{m-a-2}.$
\item For $q\geq4$, $m\geq3$, $0\leq a\leq m-3$ and $b= 3$, define for $x=(x_1,\ldots,x_m)\in\F_q^m$,
$$f(x)=\prod_{i=1}^a(1-x_i^{q-1})(x_{a+1}-u)(x_{a+2}-v)(x_{a+3}-w)$$ with $u$, $v$, $w\in\F_q$. Then, $f\in R_q(a(q-1)+3,m)$ and $|f|=(q-1)^3q^{m-a-3}>(q-2)(q-1)q^{m-a-2}$
\item For $q=3$, $m\geq4$, $1\leq a\leq m-3$ and $b=2$, define for $x=(x_1,\ldots,x_m)\in\F_q^m$, $$f(x)=\prod_{i=1}^{a-1}(1-x_i^{2})\prod_{j=1}^4(x_{a-1+j}-u_j)$$ with $u_j\in\F_q$. Then, $f\in R_3(2(a+1),m)$ and $|f|=16.3^{m-a-3}>4.3^{m-a-2}$
\end{itemize}
\end{itemize}\end{preuve}

\begin{remarque} We say that $\mathcal{B}$ is an hyperplane arrangement in $\mathcal{L}_d$ if there exist $k\in\mathbb{N}^*$, $(d_1,\ldots,d_k)\in(\mathbb{N^*})^k$ such that $\sum_{i=1}^kd_i\leq d$ and $f_1,\ldots,f_k$ are k independent linear forms over $\F_q^m$ such that $\mathcal{B}$ is composed of $k$ blocks of $d_i$ parallel hyperplanes of equation $f_i(x)=u_i,j$ where $1\leq i\leq k$, $1\leq j\leq d_i$, $u_{i,j}\in\F_q$ and if $k\neq j$, $u_{i,j}\neq u_{i,k}$. The upper bound given in the Theorem above are the third weight among hyperplane arrangements in $\mathcal{L}_d$. The proof of this result is given in Appendix.\end{remarque}

\section{Third weight}\label{poids3}

\subsection{The case where $a=0$}

We denote by $c_b$ the third weight of $R_q(b,2)$, for $2\leq b\leq q-1$. From Theorem \ref{3hyp}, we get that 
$$c_b\leq \left\{\begin{array}{ll}(q-2)(q-b+2)&\textrm{   for $q\geq 5$, $4\leq b \leq \frac{q+3}{2}$}\\&\\
(q-b+1)q &\textrm{$\begin{array}{l}\textrm{for $q\geq 7$ and $\frac{q}{2}+2 \leq b\leq q-1$ or $q\geq 3$ and $b=2$}\\ \textrm{or $q\geq 4$ and $b=3$}\end{array}$}\end{array}\right.$$

\begin{lemme}\label{t=0}Let $m\geq 2$, $q\geq 3$, $4\leq b\leq q-1$ and $f\in R_q(b,m)$. Assume $c_b<(q-b+1)q$. If $|f|>(q-b+1)(q-1)q^{m-2}$ then $|f|\geq c_bq^{m-2}$.\end{lemme}

\begin{preuve}We prove this result by induction on $m$. For $m=2$, it is the definition of $c_b$.\\

Let $m\geq 3$. Assume if $f\in R_q(b,m-1)$ is such that $|f|>(q-b+1)(q-1)q^{m-3}$ then $|f|\geq c_bq^{m-3}$.\\

Let $f\in R_q(b,m)$ such that $|f|>(q-b+1)(q-1)q^{m-2}$. Assume $|f|<c_bq^{m-2}$. We denote by $S$ the support of $f$. 

Assume $S$ meets all affine hyperplanes. Then, for all $H$ hyperplane, $\#(S\cap H)\geq (q-b)q^{m-2}$. Assume there exists $H_1$ such that $\#(S\cap H_1)=(q-b)q^{m-2}$. By applying an affine transformation, we can assume $x_1=\lambda$, $\lambda\in\F_q$ is an equation of $H_1$. We set an order on the elements of $\F_q$ such that $|f_{\lambda_1}|\leq|f_{\lambda_2}|\leq\ldots\leq |f_{\lambda_q}|$. Then, $f_{\lambda_1}$ is a minimal weight codeword of $R_q(b,m-1)$. So, by applying an affine transformation, we can assume $f_{\lambda_1}$ depends only on $x_2$. Let $k\geq 1$ be such that for all $i\leq k$, $f_{\lambda_i}$ depends only on $x_2$ and $f_{\lambda_{k+1}}$ does not depend only on $x_2$. 

If $k>b$, we can write for all $x=(x_1,\ldots,x_m)\in\F_q^m)$, $$f(x)=\sum_{i=0}^bf_{\lambda_{i+1}}^{(i)}(x_2,\ldots,x_m)\prod_{1\leq j\leq i}(x_1-\lambda_j)\qquad \textrm{(see \cite{erickson1974counting})}.$$
Since for $i \leq b+1$, $f_{\lambda_i}$ depends only on $x_2$ then $f$ depends only on $x_1$ and $x_2$ which is a contradiction by the case $m=2$. So $k\leq b$ and we can write for all $x=(x_1,\ldots,x_m)\in\F_q^m$, $$f(x)=g(x_1,x_2)+\prod_{i=1}^k(x_1-\lambda_i)h(x)$$ where $\deg(h)\leq b-k$. Then, for all $x_2\in\F_q$ and all $\underline{x}\in\F_q^{m-2}$, $$f_{\lambda_{k+1}}(x_2,\underline{x})=g_{\lambda_{k+1}}(x_2)+\alpha.h(x_2,\underline{x})$$ where $\alpha\in\F_q^*$.
So, by Lemma \ref{3.9}, since $f_{\lambda_{k+1}}$ does not depend only on $x_2$, $|f_{\lambda_{k+1}}|\geq(q-b+k)q^{m-2}$. We get \begin{align*}|f|&\geq k(q-b)q^{m-2}+(q-k)(q-b+k)q^{m-2}\\&=(q-b)q^{m-1}+(q-k)kq^{m-2}\\|f|&\geq(q-b)q^{m-1}+(q-1)q^{m-2}\end{align*}
Since $c_b<(q-b+1)q$ and $|f|<c_bq^{m-2}$, we get a contradiction.\\

Then, for all $H$ hyperplane, $\#(S\cap H)\geq (q-1)(q-b+1)q^{m-3}$. By induction hypothesis, since $|f|<c_bq ^{m-2}$ there exists an affine hyperplane $H_2$ such that $\#(S\cap H_2)=(q-1)(q-b+1)q^{m-3}$. So, there exists $A$ an affine subspace of codimension 2 included in $H_2$ which does not meet $S$ (see \cite{raey}). Then, considering all affine hyperplanes through $A$, we must have $$(q+1)(q-1)(q-b+1)q^{m-3}<c_bq^{m-2}$$ which gives, since $c_b\leq(q-b+1)q-1$, $q<q-b+1$. We get a contradiction since $b\geq 4$. \\

So there exists $H_0$ an hyperplane which does not meet $S$. We denote by $n$ the number of hyperplanes parallel to $H_0$ which do not meet $S$. By Lemma \ref{2.7}, since $c_b\leq (q-b+2)(q-2)$, we get that $n=b$, $n=b-1$ or $n=1$. By applying an affine transformation, we can assume $x_1=\lambda_1$, $\lambda_1\in\F_q$ is an equation of $H_0$.\\

If $n=b$, then for all $x=(x_1,\ldots,x_m)\in\F_q^m$, we have $$f(x)=\prod_{i=1}^b(x_1-\lambda_i)$$ with $\lambda_i\in\F_q$ and for $i\neq j$, $\lambda_i\neq \lambda_j$. In this case, $f$ is a minimum weight codeword of $R_q(b,m)$ which is absurd.\\

If $n=b-1$, then for all $x=(x_1,\ldots,x_m)\in\F_q^m$, we have $$f(x)=\prod_{i=1}^{b-1}(x_1-\lambda_i)g(x)$$ with $\lambda_i\in\F_q$, for $i\neq j$, $\lambda_i\neq \lambda_j$ and $g\in R_q(1,m)$. If $\deg(g)=0$ then $f$ is a minimum weight codeword of $R_q(b-1,m)$. If $\deg(g)=1$ then $f$ is a second weight codeword of $R_q(b,m)$. Both cases give a contradiction.\\

If $n=1$ then for all $x=(x_1,\ldots,x_m)\in\F_q^m$, we have $$f(x)=(x_1-\lambda_1)g(x)$$ with $g\in R_q(b-1,m)$. Then, for $i\geq 2$, $\deg(f_{\lambda_i})\leq (b-1)$, so, $|f_{\lambda_i}|\geq (q-b+1)q^{m-2}$. We denote by $N=\#\{i:|f_{\lambda_i}|= (q-b+1)q^{m-2}\}$. Then, 
$$N(q-b+1)q^{m-2}+(q-1-N)(q-b+2)(q-1)q^{m-3}<c_bq^{m-2}$$ which gives $N>\frac{(q-1)^2(q-b+2)-c_bq}{b-2}\geq0$ so $N\geq1$. 

Denote by $H_1$ an hyperplane such that $\#(S\cap H_1)=(q-b+1)q^{m-2}$. Then, $S\cap H_1$ is the support of a minimal weight codeword of $R_q(b-1,m-1)$ so it is the union of $(q-b+1)$ parallel affine subspaces of codimension 2 included in $H_1$. 

Now, consider $P$ an affine subspace of codimension 2 included in $H_1$ such that $\#(S\cap P)=(q-b+1)q^{m-3}$. Then, for all $H$ hyperplane through $P$, $\#(S \cap H)\geq (q-b+1)(q-1)q^{m-3}$. Indeed, by definition of $P$, $S$ meets all hyperplanes through $P$, so, for all $H$ hyperplane through $P$, $\#(S\cap H)\geq (q-b)q^{m-2}$. If $\#(S\cap H)= (q-b)q^{m-2}$, then $S\cap H$ is the union of $(q-b)$ parallel affine subspaces of codimension 2 which is absurd since it intersects $P$ in $(q-b+1)$ affine subspaces of codimension 3. We can apply the same argument to all affine subspaces of codimension 2 included in $H_1$ parallel to $P$. Now consider an hyperplane through $P$ and the $q$ hyperplanes parallel to this hyperplane, since $|f|<c_bq^{m-2}$, one of these hyperplanes, say $H_2$, meets $S$ in $(q-b+1)(q-1)q^{m-3}$ points.

We denote by $(A_i)_{1\leq i\leq b}$ the $b$ affine subspaces of codimension 2 included in $H_2$ which do not meet $S$. Let $1\leq i\leq b$, suppose that $S$ meets all hyperplanes through $A_i$ and let $H$ be one hyperplane through $A_i$. If all hyperplanes parallel to $H$ meet $S$ then as in the beginning of the proof of this lemma, we get $\#(S\cap H)\geq(q-1)(q-b+1)q^{m-3}$. If there exists an hyperplane parallel to $H$ which does not meet $S$ then $\#(S\cap H)\geq(q-b+1)q^{m-2}$. In both cases we get a contradiction since $(q+1)(q-b+1)(q-1)q^{m-3}\geq c_bq^{m-2}$. So, for all $1\leq i\leq b$ there exists an hyperplane through $A_i$ which does not meet $S$.

Then  at least $b-1$  of the hyperplanes through the $(A_i)$ which do not meet $S$ must intersect $H_1$, we get that $|f|=(q-1)(q-b+1)q^{m-2}$ (see \cite{raey}) which is absurd.

\begin{figure}[!h]
\caption{}
\begin{center}\subfloat[]{\label{fig1a}
\begin{tikzpicture}[scale=0.17]
\draw (0,0)--(5,2)--(5,11)--(0,9)--cycle;
\draw (1,2/5)--(1,9+2/5);
\draw (2,4/5)--(2,9+4/5);
\draw[dotted] (3,6/5)--(3,9+6/5);
\draw[dotted] (4,8/5)--(4,9+8/5);
\draw (0,9) node[above left]{$H_1$};
\draw[dashed](0,4)--++(5,2);
\draw (0,4)--(12,4)--++(5,2)--++(-12,0);
\draw (17,6) node[above right]{$H_2$};
\draw (1,4+2/5)--++(12,0);
\draw (2, 4+4/5)--++(12,0);
\draw[dotted] (3,4+6/5)--++(12,0);
\draw[dotted] (4,4+8/5)--++(12,0);
\draw (4,8/5)--++(12,0)--++(0,9)--++(-12,0);
\draw[dashed] (4,4+8/5)--(11,4);
\draw (0,4) node[left]{$P$};
\end{tikzpicture}}
\hspace{3cm}
\subfloat[]{\label{fig1b}
\begin{tikzpicture}[scale=0.17]
\draw (0,0)--(5,2)--(5,11)--(0,9)--cycle;
\draw (1,2/5)--(1,9+2/5);
\draw (2,4/5)--(2,9+4/5);
\draw[dotted] (3,6/5)--(3,9+6/5);
\draw[dotted] (4,8/5)--(4,9+8/5);
\draw (0,9) node[above left]{$H_1$};
\draw[dashed](0,4)--++(5,2);
\draw (0,4)--(12,4)--++(5,2)--++(-12,0);
\draw (17,6) node[above right]{$H_{2}$};
\draw (2,4+4/5)--++(12,0);
\draw (1,4+2/5)--(15,4+6/5);
\draw[dotted] (3,4+6/5)--(13,4+2/5);
\draw[dotted] (4,4+8/5)--(12,4);
\draw[dotted] (6,4)--(11,6);
\draw (6,0)--++(5,2)--++(0,9)--++(-5,-2)--cycle;
\draw (0,4) node[left]{$P$};
\end{tikzpicture}}\end{center}
\label{fig1}
\end{figure}
\end{preuve}

\begin{lemme}Let $m\geq2$, $q\geq 3$ and $f\in R_q(2,m)$. If $|f|>(q-1)^2q^{m-2}$ then $|f|\geq (q^2-q-1)q^{m-2}$.\end{lemme}

\begin{preuve}Let $f\in R_q(2,m)$ such that $|f|>(q-1)^2q^{m-2}$. If $\deg(f)\leq 1$ then $|f|\geq(q-1)q^{m-1}$. From now, assume that $\deg(f)=2$.\\
First we recall some results on quadratic forms. These results can be found in \cite{hirschfeld_proj_geom} for example. 
If $Q$ is a quadratic form of rank $R$ on $\F_q^m$ then, there exists a linear transformation such that for all $x=(x_1,\ldots,x_m)\in\F_q^m$, 
\\if $R=2r+1$ \begin{equation}\label{eq1}Q(x)=\sum_{i=1}^rx_{2i-1}x_{2i}+ax_{2r+1}^2\end{equation}
or if $R=2r$
\begin{equation}\label{eq2}Q(x)=\sum_{i=1}^rx_{2i-1}x_{2i}\end{equation}
or \begin{equation}\label{eq3}Q(x)=\sum_{i=1}^{r-1}x_{2i-1}x_{2i}+ax_{2r-1}^2+bx_{2r-1}x_{2r}+cx_{2r}^2\end{equation} with $ax^2+bx+c$ is irreducible over $\F_q$.

Then $N(Q)$ the number of zeros of $Q$ is $$N(Q)=q^{m-1}+(w-1)(q-1)q^{m-\frac{R}{2}-1}$$ 
where $$w=\left\{\begin{array}{ll}1&\textrm{if $R$ is odd}\\2&\textrm{if $R$ is even and $Q$ is of type \eqref{eq2}}\\0&\textrm{if $R$ is even and $Q$ is of type \eqref{eq3}}\end{array}\right..$$ 
We write for all $x=(x_1,\ldots,x_m)\in\F_q^m$, $f(x)=q_0(x)+\alpha_1x_1+\ldots+ \alpha_mx_m+\beta$ where $q_0$ is a quadratic form of rank $r$ and $w_0$ is defined as above. Then the number of zeros of $f$ is the number of affine zeros of the homogeneized form $Q(x)=q_0(x)+\alpha_1x_1z+\ldots+ \alpha_mx_mz+\beta z^2$. We denote by $R$ the rank of $Q$ and $w$ is defined as above. Then, using the formula above, $$|f|=(q-1)q^{m-1}+(w_0-1)q^{m-\frac{r}{2}-1}-(w-1)q^{m-\frac{R}{2}}.$$. By applying affine transformation (see \cite{6362214}), we get that :
\begin{itemize}
\item If $r$ is odd then, either $R=r$, $w=1$ and $|f|=(q-1)q^{m-1}$ or $R=r+1$, $w=2$ and $|f|=(q-1)q^{m-1}-q^{m-\frac{r+1}{2}}$.
\item If $r$ is even and $w_0=2$ then, $R=r+2$, $w=2$ and $|f|=(q-1)q^{m-1}$, $R=r$, $w=2$ and $|f|=(q-1)(q^{m-1}-q^{m-1-\frac{r}{2}})$ or $R=r+1$, $w=1$ and $|f|=(q-1)q^{m-1}+q^{m-1-\frac{r}{2}}$.
\item If $r$ is even and $w_0=0$ then, $R=r+2$, $w=0$ and $|f|=(q-1)q^{m-1}$, $R=r+1$, $w=1$ and $|f|=(q-1)q^{m-1}-q^{m-1-\frac{r}{2}}$ or $R=r$, $w=0$ and $|f|=(q-1)(q^{m-1}+q^{m-1-\frac{r}{2}})$.
\end{itemize}
Finally, the third weight of $R_q(2,m)$ is $(q^2-q-1)q^{m-2}$.
\end{preuve}

\begin{lemme}For $q\geq4$, $c_3=q^2-3q+3$. 

Furthermore, for $q\geq 7$, if $f\in R_q(3,2)$ is such that $|f|=q^2-3q+3$ then up to affine transformation for all $(x,y)\in\F_q^2$,
$$f(x,y)=(a_1x+b_1y)(a_2x+b_2y)(a_3x+b_3y+c)$$ where $(a_i,b_i)\in\F_q^2\setminus\{(0,0)\}$ such that for $i\neq j$, $a_ib_j-a_jb_i\neq0$ and $c\in\F_q^*$.\end{lemme}

\begin{preuve}The second weight in this case is $(q-2)(q-1)=q^2-3q+2$. So we only have to find a codeword of $R_q(3,2)$ such that its weight is $q^2-3q+3$ to prove the first part of this proposition. Consider $3$ lines which meet pairwise but do not intersect in one point. Then the union of this 3 lines has $3q-3$ points. Let $a_1x+b_1y+c_1=0$, $a_2x+b_2y+c_2=0$ and $a_3x+b_3y+c_3=0$ be the equations of these 3 lines then $f(x,y)=\displaystyle\prod_{i=1}^3(a_ix+b_iy+c_i)\in R_q(3,2)$ and $|f|=q^2-3q+3$.

Let $f\in R_q(3,2)$ such that $|f|=q^2-3q+3$. Denote by $S$ the support of $f$. For $q\geq4$, $q^2-3q+3<(q-2)q$. Since $(q-2)q$ is the minimum weight of $R_q(2,2)$, $\deg(f)=3$. We prove first that for $q\geq7$, $f$ is the product of $3$ affine factors. 
Let $P$ be a point of $\F_q^2$ which is not in $S$ and $L$ be a line in $\F_q^2$ such that $P\in L$. Then, either $L$ does not meet $S$ or $L$ meets $S$ in at least $q-3$ points. If any line through $P$ meets $S$ then $$(q+1)(q-3)\leq|f|\leq q^2-3q+3$$ which is absurd since $q\geq7$. So there exists a line through $P$ which does not meet $S$. By applying the same argument to all $P$ not in $S$, we get that $f$ is the product of affine factors.

Denote by $Z$ the set of zeros of $f$. We have just proved that this set is the union of 3 lines. If these 3 lines are parallel then we get a minimum weight codeword. If 2 of these lines are parallel or the 3 lines meet in one point, we have a second weight codeword. So the only possibility is the case where the 3 lines meet pairwise but do not intersect in one point which gives the result.\end{preuve}

\begin{lemme}\label{m=3}Let $q\geq 4$. If $f\in R_q(3,3)$ and $|f|>(q-1)(q-2)q$ then $|f|\geq (q-1)^3$.\end{lemme}

\begin{preuve} Let $f\in R_q(3,3)$ such that $|f|>(q-2)(q-1)q$. Assume $|f|<(q-1)^3$. We denote by $S$ the support of $f$. 

Assume $S$ meets all hyperplanes. Then for all $H$ hyperplane, $\#(S\cap H)\geq(q-3)q$. Assume there exists $H_1$ an hyperplane such that $\#(S\cap H_1)=(q-3)q$. By applying an affine transformation, we can assume $x_1=0$ is an equation of $H_1$.
We set an order on the elements of $\F_q$ such that $|f_{\lambda_1}|\leq\ldots\leq|f_{\lambda_q}|$. Since $f_{\lambda_1}$ is a minimum weight codeword of $R_q(3,2)$, by applying an affine transformation, we can assume it depends only on $x_2$. Let $k\geq1$ be such that for all $i\leq k$, $f_{\lambda_i}$ depends only on $x_2$ but $f_{\lambda_{k+1}}$ does not depend only on $x_2$. If $k\geq 3$ then we can write for all $(x_1,x_2,x_3)\in\F_q$ , $$f(x_1,x_2,x_3)=g(x_1,x_2)+(x_1-\lambda_1)(x_1-\lambda_2)(x_1-\lambda_3)h(x_1,x_2,x_3)$$ where $\deg(h)\leq 3-3=0$ and $f$ depends only on $x_1$ and $x_2$. So, $|f|\equiv0\mod q$. But since $|f|>(q-1)(q-2)q$, $|f|\geq (q-1)(q-2)q+q\geq (q-1)^3$ which gives a contradiction . So, $k\leq 2$. Since $f_{\lambda_1},\ldots,f_{\lambda_k}$ depend only on $x_2$, we can write for all $(x_1,x_2,x_3)\in\F_q^3$, $$f(x_1,x_2,x_3)=g(x_1,x_2)+(x_1-\lambda_1)\ldots(x_1-\lambda_k)h(x_1,x_2,x_3)$$ where $\deg(h)\leq 3-k$. 
Then, $$f_{\lambda_{k+1}}(x_2,x_3)=g_{\lambda_{k+1}}(x_2)+\alpha h_{\lambda_{k+1}}(x_2,x_3)$$ where $\alpha\in \F_q^*$. So by Lemma \ref{3.9}, since $f_{\lambda_{k+1}}$ does not depends only on $x_2$, $|f_{\lambda_{k+1}}|\geq(q-3+k)q$.
We get $$|f|\geq k(q-3)q+(q-k)(q-3+k)q=(q-3)q^2+(q-k)kq\geq(q-3)q^2+(q-1)q\geq (q-1)^3$$ which is absurd since $q\geq4$.

So for all $H$ hyperplane, $\#(S\cap H)\geq (q-1)(q-2)$. Considering $q$ parallel hyperplanes, since $((q-1)(q-2)+1)q\geq(q-1)^3$, there exits an hyperplane $H_0$ such that $\#(S\cap H_0)= (q-1)(q-2)$. 

So there exists $A$ an affine subspace of codimension 2 included in $H_0$ which does not meet $S$. Considering all hyperplanes through $A$, since $S$ meets all hyperplanes, we get 
$$(q+1)(q-1)(q-2)\leq |f|<(q-1)^3$$ which is absurd since $q\geq4$. So there exists $H_1$ an affine hyperplane which does not meet $S$. We denote by $n$ the number of hyperplanes parallel to $H_1$ which do not meet $S$. 

By applying an affine transformation, we can assume $x_1=\lambda_1$ is an equation of $H_1$. by Lemma \ref{DGMW1}, $n\leq 3$

If $n=3$, then by Lemma \ref{DGMW1}, we have for all $x=(x_1,x_2,x_3)\in\F_q^3$ $$f(x)=(x_1-\lambda_1)(x_1-\lambda_2)(x_1-\lambda_3)g(x)$$ where $\lambda_i\in\F_q$, $\lambda_i\neq\lambda_j$ for $i\neq j$, $\deg(g)\leq 0$. So, $|f|=(q-3)q^2$ which is absurd.

If $n=2$, then by Lemma \ref{DGMW1}, we have for all $x=(x_1,x_2,x_3)\in\F_q^3$ $$f(x)=(x_1-\lambda_1)(x_1-\lambda_{2})g(x)$$ where $\lambda_2\in\F_q$, $\lambda_2\neq\lambda_1$, $\deg(g)\leq 1$. So, if $\deg(g)=0$, $|f|=(q-2)q^2$. If $\deg(g)=1$, $|f|=(q-2)(q-1)q$. Both cases give a contradiction.

If $n=1$, then by Lemma \ref{DGMW1}, we have for all $x=(x_1,x_2,x_3)\in\F_q^3$ $$f(x)=(x_1-\lambda_1)g(x)$$ where $\deg(g)\leq 2$. Then, for $i\geq 2$, $\deg(f_{\lambda_i})\leq 2$, so, $|f_{\lambda_i}|\geq (q-2)q$. We denote by $N=\#\{i:|f_{\lambda_i}|= (q-2)q\}$. Then, 
$$N(q-2)q+(q-1-N)(q-1)^2\leq|f|<(q-1)^3$$ so $N\geq1$. 

Denote by $H_2$ an hyperplane such that $\#(S\cap H_2)=(q-2)q$. Then, $S\cap H_2$ is the support of a minimal weight codeword of $R_q(2,2)$ so it is the union of $(q-2)$ parallel affine subspaces of codimension 2 included in $H_2$. 
Now, consider $P$ an affine subspace of codimension 2 included in $H_2$ such that $\#(S\cap P)=(q-2)$. Then, for all $H$ hyperplane through $P$, $\#(S \cap H)\geq (q-2)(q-1)$. Indeed, by definition of $P$, $S$ meets all hyperplanes through $P$, so, for all $H$ hyperplane through $P$, $\#(S\cap H)\geq (q-3)q$. If $\#(S\cap H)= (q-3)q$, then $S\cap H$ is the union of $(q-3)$ parallel affine subspaces of codimension 2 which is absurd since it intersects $P$ in $(q-2)$ affine subspaces of codimension 3. We can apply the same argument to all affine subspaces of codimension 2 included in $H_2$ parallel to $P$. Now consider an hyperplane through $P$ and the $q$ hyperplanes parallel to this hyperplane, since $|f|<(q-1)^3$, one of these hyperplanes, say $H_3$, meets $S$ in $(q-2)(q-1)$ points. 

We denote by $(A_i)_{1\leq i\leq 3}$ the $3$ affine subspaces of codimension 2 included in $H_3$ which do not meet $S$. Suppose that $S$ meets all hyperplanes through $A_i$ and consider $H$ one of them. If all hyperplanes parallel to $H$ meet $S$ then as in the beginning of the proof of this lemma, we get that $\#(S\cap H)\geq (q-1)(q-2)$. If there exists an hyperplane parallel to $H$ which does not meet $S$ then $\#(S\cap H)\geq (q-2)q$. In all cases we get a contradiction since $(q+1)(q-1)(q-2)\geq(q-1)^3$.

Then  at least $2$  of the hyperplanes through the $(A_i)$ which do not meet $S$ must intersect $H_2$, we get that $|f|=(q-1)(q-b+1)q$ (see \cite{raey}) which is absurd.\end{preuve}

\begin{lemme}\label{t=03}Let $q\geq 4$ and $m \geq3$. If $f\in R_q(3,m)$ and $|f|>(q-1)(q-2)q^{m-2}$ then, $|f|\geq(q-1)^3q^{m-3}$.\end{lemme}

\begin{preuve}We prove this lemma by induction on $m$. The case where $m=3$ comes from lemma \ref{m=3}.
Assume for some $m\geq4$, if $f\in R_q(3,m-1)$ is such that $|f|>(q-1)(q-2)q^{m-3}$, then $|f|\geq (q-1)^3q^{m-4}$.

Let $f\in R_q(3,m)$ such that $|f|>(q-2)(q-1)q^{m-2}$. Assume $|f|<(q-1)^3q^{m-3}$. We denote by $S$ the support of $f$. 

Assume $S$ meets all hyperplanes. Then for all $H$ hyperplane, $\#(S\cap H)\geq(q-3)q^{m-2}$. Assume there exists $H_1$ such that $\#(S\cap H_1)=(q-3)q^{m-2}$. By applying an affine transformation, we can assume $x_1=0$ is an equation of $H_1$.
We set an order on the elements of $\F_q$ such that $|f_{\lambda_1}|\leq\ldots\leq|f_{\lambda_q}|$. Since $f_{\lambda_1}$ is a minimum weight codeword of $R_q(3,m-1)$, by applying an affine transformation, we can assume it depends only on $x_2$. Let $k\geq1$ be such that for all $i\leq k$, $f_{\lambda_i}$ depends only on $x_2$ but $f_{\lambda_{k+1}}$ does not depend only on $x_2$. If $k\geq 3$ then we can write for all $x_1$ , $x_2\in\F_q$ and $\underline{x}\in\F_q^{m-2}$, $$f(x_1,x_2,\underline{x})=g(x_1,x_2)+(x_1-\lambda_1)\ldots(x_1-\lambda_3)h(x_1,x_2,\underline{x})$$ where $\deg(h)\leq 0$. So, $f$ depends only on $x_1$ and $x_2$ and $|f|\equiv 0\mod q^{m-2}$. Since $|f|>(q-1)(q-2)q^{m-2}$, $|f|\geq(q-1)(q-2)q^{m-2}+q^{m-2}\geq(q-1)^3q^{m-3}$ which gives a contradiction. So, $k\leq 2$. Since $f_{\lambda_1},\ldots,f_{\lambda_k}$ depend only on $x_2$ we can write for all $x_1$ , $x_2\in\F_q$ and $\underline{x}\in\F_q^{m-2}$, $$f(x_1,x_2,\underline{x})=g(x_1,x_2)+(x_1-\lambda_1)\ldots(x_1-\lambda_k)h(x_1,x_2,\underline{x})$$ where $\deg(h)\leq 3-k$. 
Then $$f_{\lambda_{k+1}}(x_2,\underline{x})=g_{\lambda_{k+1}}(x_2)+\alpha h_{\lambda_{k+1}}(x_2,\underline{x})$$ where $\alpha\in \F_q^*$. So by Lemma \ref{3.9}, since $f_{\lambda_{k+1}}$ does not depend only on $x_2$, $|f_{\lambda_{k+1}}|\geq(q-3+k)q^{m-2}$.
We get $$|f|\geq k(q-3)q^{m-2}+(q-k)(q-3+k)q^{m-2}=(q-3)q^{m-1}+(q-k)kq^{m-2}.$$
Since $|f|<(q-1)^3q^{m-3}$, this is absurd.

So for all $H$ hyperplane, $\#(S\cap H)\geq (q-1)(q-2)q^{m-3}$. By induction hypothesis, considering $q$ parallel hyperplanes, there exits an hyperplane $H_0$ such that $\#(S\cap H_0)= (q-1)(q-2)q^{m-3}$. 

So there exists $A$ an affine subspace of codimension 2 included in $H_0$ which does not meet $S$. Considering all hyperplanes through $A$, since $S$ meets all hyperplanes, we get 
$$(q+1)(q-1)(q-2)q^{m-2}<(q-1)^3q^{m-3}$$ which is absurd. So there exists an affine hyperplane $H_1$ which does not meet $S$. We denote by $n$ the number of hyperplanes parallel to $H_1$ which do not meet $S$.  

By applying an affine transformation, we can assume $x_1=\lambda_1$ is an equation of $H_1$. By Lemma \ref{DGMW1}, $n\leq3$.

If $n=3$, then by Lemma \ref{DGMW1}, we have for all $x=(x_1,\ldots,x_m)\in\F_q^m$ $$f(x)=(x_1-\lambda_1)(x_1-\lambda_2)(x_1-\lambda_3)g(x)$$ where $\lambda_i\in\F_q$, $\lambda_i\neq\lambda_j$ for $i\neq j$, $\deg(g)\leq 0$. So, $|f|=(q-3)q^{m-1}$ which is absurd.

If $n=2$, then by Lemma \ref{DGMW1}, we have for all $x=(x_1,\ldots,x_m)\in\F_q^m$ $$f(x)=(x_1-\lambda_1)(x_1-\lambda_{2})g(x)$$ where $\lambda_2\in\F_q$, $\lambda_2\neq\lambda_1$, $\deg(g)\leq 1$. So, if $\deg(g)=0$, $|f|=(q-2)q^{m-1}$. If $\deg(g)=1$, $|f|=(q-2)(q-1)q^{m-2}$. Both cases give a contradiction.

If $n=1$, then by Lemma \ref{DGMW1}, we have for all $x=(x_1,\ldots,x_m)\in\F_q^m$ $$f(x)=(x_1-\lambda_1)g(x)$$ where $\deg(g)\leq2$. Then, for $i\geq 2$, $\deg(f_{\lambda_i})\leq 2$, so, $|f_{\lambda_i}|\geq (q-2)q^{m-2}$. We denote by $N=\#\{i:|f_{\lambda_i}|= (q-2)q^{m-2}\}$. Then, 
$$N(q-2)q^{m-2}+(q-1-N)(q-1)^2q^{m-3}\leq|f|<(q-1)^3q^{m-3}$$ which gives $N\geq1$. 

Denote by $H_2$ an hyperplane such that $\#(S\cap H_2)=(q-2)q^{m-2}$. Then, $S\cap H_2$ is the support of a minimal weight codeword of $R_q(2,m-1)$ so it the union of $(q-2)$ parallel affine subspaces of codimension 2 included in $H_2$.

Now, consider $P$ an affine subspace of codimension 2 included in $H_2$ such that $\#(S\cap P)=(q-2)q^{m-3}$. Then, for all $H$ hyperplane through $P$, $\#(S \cap H)\geq (q-2)(q-1)q^{m-3}$. Indeed, by definition of $P$, $S$ meets all hyperplane through $P$, so, for all $H$ hyperplane through $P$, $\#(S\cap H)\geq (q-3)q^{m-2}$. If $\#(S\cap H)= (q-3)q^{m-2}$, then $S\cap H$ is the union of $(q-3)$ parallel affine subspaces of codimension 2 which is absurd since it intersects $P$ in $(q-2)$ affine subspaces of codimension 3.  We can apply the same argument to all affine subspaces of codimension 2 included in $H_2$ parallel to $P$. Now consider an hyperplane through $P$ and the $q$ hyperplanes parallel to this hyperplane, since $|f|<(q-1)^3q^{m-3}$, one of these hyperplanes, say $H_3$, meets $S$ in $(q-2)(q-1)q^{m-3}$ points.

We denote by $(A_i)_{1\leq i\leq 3}$ the $3$ affine subspaces of codimension 2 included in $H_3$ which do not meet $S$. Suppose that $S$ meets all hyperplanes through $A_i$ and consider $H$ one of them. If all hyperplanes parallel to $H$ meet $S$ then as in the beginning of the proof of this lemma, we get that $\#(S\cap H)\geq (q-1)(q-2)q^{m-3}$. If there exists an hyperplane parallel to $H$ which does not meet $S$ then $\#(S\cap H)\geq (q-2)q^{m-2}$. In all cases we get a contradiction since $(q+1)(q-1)(q-2)q^{m-3}\geq(q-1)^3q^{m-3}$.

Then  at least $2$  of the hyperplanes through the $(A_i)$ which do not meet $S$ must intersect $H_2$, we get that $|f|=(q-1)(q-b+1)q^{m-2}$ (see \cite{raey}) which is absurd.\end{preuve}

\subsection{The case where $a$ is maximal}

\begin{lemme}\label{m-2}Let $m\geq 4$, $q\geq5$ and $2\leq b\leq q-2$. Assume $c_b<(q-b+1)q$. If $f\in R_q((m-2)(q-1)+b,m)$ and $|f|>(q-1)(q-b+1)$, then $|f|\geq c_b$.\end{lemme}

\begin{preuve}Let $f\in R_q((m-2)(q-1)+b,m)$ such that $|f|>(q-b+1)(q-1)$. Assume $|f|<c_b$ and denote by $S$ the support of $f$. 

Assume $S$ meets all affine hyperplanes. We set an order on the elements of $\F_q$ such that $|f_{\lambda_1}|\leq\ldots\leq|f_{\lambda_q}|$. Then for all $H$ hyperplane, $\#(S\cap H)\geq q-b$ and since $(q-b+1)q>c_b$, $|f_{\lambda_1}=(q-b)$. By applying an affine transformation, we can assume $(1-x_2^{q-1})$ divides $f_{\lambda_1}$. Let $1\leq k$ be such that for all $i\leq k$, $(1-x_2^{q-1})$ divides $f_{\lambda_i}$ and $(1-x_2^{q-1})$ does not divide $f_{\lambda_{k+1}}$. Then, by Lemma \ref{3.7}, $|f|\geq (q-b)q+(q-1)$. We get a contradiction since $ (q-b)q+(q-1)\geq c_b$.

So there exists an hyperplane $H_0$ which does not meet $S$. By applying an affine transformation, we can assume $x_1=\alpha$, $\alpha\in\F_q$, is an equation of $H_0$. We denote by $n$ the number of hyperplanes parallel to $H_0$ which do not meet $S$. We set an order on the elements of $\F_q$ such that $|f_{\lambda_1}|\leq\ldots\leq|f_{\lambda_q}|$.

If $n=q-1$ then we can write for all $x=(x_1,\ldots,x_m)\in\F_q^m$, $$f(x)=(1-x_1^{q-1})g(x_2,\ldots,x_m)$$ where $g\in R_q((m-3)(q-1)+b,m-1)$ and $|f|=|g|$. So, $g$ fulfils the same conditions as $f$ with one variable less. Iterating this process, we end either in the case where $t=0$ (which is absurd by definition of $c_b$) or in the case where $n<q-1$.

From now, we assume $n<(q-1)$. By Lemma \ref{2.7}, since for $b\geq 3$, $|f|<c_b\leq (q-b+2)(q-2)$ and for $b=2$, $q^2-q-1\leq 2(q-3)q$, the only possibilities for $b\geq 2$ are $n=1$, $n=b-1$ or $n=b$. We can write for all $x=(x_1,\ldots,x_m)\in\F_q^m$ $$f(x)=\prod_{1\leq i\leq n}(x_1-\lambda_i)g(x)$$ where $g\in R_q((m-2)(q-1)+b-n,m)$. Then for all $i\geq n+1$, $f_{\lambda_i}\in R_q((m-2)(q-1)+b-n,m-1)$  and $|g_{\lambda_i}|=|f_{\lambda_i}|\geq (q-b+n)$.

Assume $n=b$. For $\lambda\in\F_q$, if $|g_{\lambda}|>q$, then $|g_{\lambda}|\geq 2(q-1)$. Denote by $N:=\#\{i\geq b+1 :|g_{\lambda_i}|=q\}$. Since for $i\geq b+1$, $|f_{\lambda_i}|=|g_{\lambda_i}|$ and $(q-b)2(q-1)\geq(q-b+1)q$ for $b\leq q-2$, we get $N\geq1$. Furthermore, since $(q-b)q<(q-b+1)(q-1)<|f|$, $N\leq q-b-1$. 

Assume $|f_{\lambda_{b+N+1}}|\geq(N+1)q$. Then $$Nq+(q-b-N)(N+1)q\leq |f|<c_b$$ which gives $$Nq(q-N-b)< c_b-(q-b)q<q.$$ This is absurd since $1\leq N\leq q-b-1$. Furthermore the only possibility such that $|f_{\lambda_{b+N+1}}|=Nq$ is $N=1$ which is absurd since $f_{\lambda_{b+N+1}}$ is not a minimal weight codeword.

By Lemma \ref{2.15.1}, for all $b+1\leq i\leq N+b$, $g_{\lambda_{b+1}}=g_{\lambda_i}$. So, we can write for all $x=(x_1,\ldots,x_m)\in\F_q^m$ \begin{align*}f(x)&=\prod_{1\leq i\leq b}(x_1-\lambda_i)\left(g_{\lambda_{b+1}}(x_2,\ldots,x_m)+ \prod_{b+1\leq i\leq N+b}(x_1-\lambda_i)h(x)\right)\\&=\prod_{1\leq i\leq b}(x_1-\lambda_i)\left(\alpha f_{\lambda_{b+1}}(x_2,\ldots,x_m)+ \prod_{b+1\leq i\leq N+b}(x_1-\lambda_i)h(x)\right)\end{align*} where $h\in R_q((m-2)(q-1)-N,m)$ and $\alpha\in\F_q^*$. 

Then, for all $(x_2,\ldots,x_m)\in\F_q^{m-1}$, $$f_{\lambda_{b+N+1}}(x_2,\ldots,x_m)=\beta f_{\lambda_{b+1}}(x_2,\ldots,x_m)+\gamma h_{\lambda_{b+N+1} }(x_2,\ldots,x_m).$$ We get a contradiction by Lemma \ref{2.14}.

Now, assume $n=1$ or $n=b-1$.

If $(1-x_2^{q-1})$ divides $f$, we can write for all $x=(x_1,\ldots,x_m)\in\F_q^m$, $$f(x)=(1-x_2^{q-1})h(x_1,x_3,\ldots,x_m)$$ where $h\in R_q((m-3)(q-1)+b,m-1)$ and $|f|=|h|$. So, $h$ fulfils the same conditions as $f$. Iterating this process, we end either in the case where $t=0$ which is absurd by definition of $c_b$ or in the case where $(1-x_2^{q-1})$ does not divide $h$. So we can assume $(1-x_2^{q-1})$ does not divide $f$. 

Since $n\geq 1$, $f_{\lambda_1}=0$. So, $1-x_2^{q-1}$ divides $f_{\lambda_1}$. Then, since $1-x_2^{q-1}$ does not divide $f$, there exists $k\in\{1,\ldots,q-1\}$ such that for all $i\leq k$, $1-x_2^{q-1}$ divides $f_{\lambda_i}$ and $(1-x_2^{q-1})$ does not divide $f_{\lambda_{k+1}}$. For $\lambda\in\F_q$, if $|f_{\lambda}|>(q-b+n)$ then $$|f_{\lambda}|\geq w_2=\left\{\begin{array}{ll}q&\textrm{if $n=b-1$}\\(q-b+2)&\textrm{if $n=1$}\end{array}\right..$$ Denote by $N:=\#\{i\geq n+1 :|f_{\lambda_i}|=(q-b+n)\}$. In all cases, $(q-n)w_2\geq c_b.$ So, $N\geq1$. Furthermore $(q-b+n)(q-n)=(q-b+1)(q-1)<|f|$ so $N\leq q-n-1$. 

Then, $|f_{\lambda_{n+1}}|=(q-b+n)$ and $f_{\lambda_{n+1}}$ is a minimal weight codeword of $R_q((m-2)(q-1)+b-n,m-1)$ so, by applying an affine transformation, we can assume $1-x_2^{q-1}$ divides $f_{\lambda_{n+1}}$. Thus, $k\geq n+1\geq 2$.

If $1\leq k\leq n+N-1$, then $|f_{\lambda_{k+1}}|=(q-b+n)<(q-b+k)$. If $n+N\leq k\leq q-1$, assume $|f_{\lambda_{k+1}}|\geq (q-b+k)$. Then, $$|f|\geq N(q-b+n)+(k-n-N)w_2+(q-k)(q-b+k)\geq (q-b+1)q-1$$ for $b\geq 4$, $n+N\leq k\leq q-1$ and $1\leq N\leq q-n-1$. So, we get a contradiction since $|f|<c_b<(q-b+1)q$.

Since for all $n\leq i\leq k$, $1-x_2^{q-1}$ divides $f_{\lambda_i}$, it divides $g_{\lambda_i}$ too. Then we can write for all $x=(x_1,x_2,\ldots,x_m)\in\F_q^{m}$ \begin{align*}f(x)&=\prod_{1\leq i\leq n}(x_1-\lambda_i)\left(\prod_{n+1\leq i\leq k}(x_1-\lambda_i)h(x_1,x_2,x_3,\ldots,x_m)\right.\\&\hspace{2cm}+(1-x_2^{q-1})l(x_1,x_3,\ldots,x_m)\Bigg)\end{align*} with $\deg(h)\leq (m-2)(q-1)+b-k$. Then for all $(x_2,\ldots,x_m)\in\F_q^{m-1}$, $$f_{\lambda_{k+1}}(x_2, \ldots,x_m)=\alpha h_{\lambda_{k+1}}(x_2,\ldots,x_m)+\beta(1-x_2^{q-1})l_{\lambda_{k+1}}(x_3,\ldots,x_m).$$ 
 Thus, we get a contradiction by Lemma \ref{3.5} since $k\geq2$ and $|f_{\lambda_{k+1}}|<(q-b+k)$.
\end{preuve}

\begin{lemme}\label{m-3}Let $m\geq 4$, $q\geq 7$. If $f\in R_q((m-3)(q-1)+3,m)$ and $|f|>(q-1)(q-2)q$ then $|f|\geq (q-1)^3$.\end{lemme}

\begin{preuve}Let $f\in R_q((m-3)(q-1)+3,m)$ such that $|f|>(q-2)(q-1)q$. Assume $|f|<(q-1)^3$ and denote by $S$ the support of $f$. 

Assume $S$ meets all affine hyperplanes. We set an order on the elements of $\F_q$ such that $|f_{\lambda_1}|\leq\ldots\leq|f_{\lambda_q}|$. Then for all $H$ hyperplane, $\#(S\cap H)= (q-3)q$ or $\#(S\cap H)\geq (q-2)(q-1)$ and since $((q-2)(q-1)+1)q\geq (q-1)^3$, $|f_{\lambda_1}|\leq(q-2)(q-1)$. By applying an affine transformation, we can assume $(1-x_2^{q-1})$ divides $f_{\lambda_1}$. Let $1\leq k$ be such that for all $i\leq k$, $(1-x_2^{q-1})$ divides $f_{\lambda_i}$ but $(1-x_2^{q-1})$ does bot divide $f_{\lambda_{k+1}}$ $f_{\lambda_{k+1}}$. Then, by Lemma \ref{3.7}, $|f|\geq (q-3)q^2+(q-1)q$. We get a contradiction since $ (q-3)q^2+(q-1)q\geq (q-1)^3$.

So there exists an hyperplane $H_0$ which does not meet $S$. By applying an affine transformation, we can assume $x_1=\alpha$, $\alpha\in\F_q$, is an equation of $H_0$. We denote by $n$ the number of hyperplanes parallel to $H_0$ which do not meet $S$. We set an order on the elements of $\F_q$ such that $|f_{\lambda_1}|\leq\ldots\leq|f_{\lambda_q}|$.

If $n=q-1$ then, we can write for all $x=(x_1,\ldots,x_m)\in\F_q^m$, $$f(x)=(1-x_1^{q-1})g(x_2,\ldots,x_m)$$ where $g\in R_q((m-4)(q-1)+3,m-1)$ and $|f|=|g|$. So, $g$ fulfils the same conditions as $f$ with one variable less. Iterating this process, we end either in the case where $t=0$ (which gives a contradiction) or in the case where $n<q-1$.

From now, we assume $n<(q-1)$. By Lemma \ref{2.7}, since $(q-1)^3\leq 2(q-4)q^2$, the only possibilities are $n=1$, $n=2$
or $n=3$. We can write for all $x=(x_1,\ldots,x_m)\in\F_q^m$ $$f(x)=\prod_{1\leq i\leq n}(x_1-\lambda_i)g(x)$$ where $g\in R_q((m-3)(q-1)+3-n,m)$. Then for all $i\geq n+1$, $f_{\lambda_i}\in R_q((m-3)(q-1)+3-n,m)$  and $|f_{\lambda_i}|=|g_{\lambda_i}|\geq (q-3+n)q$.

Assume $n=3$. For $\lambda\in\F_q$, if $|g_{\lambda}|>q^2$, then $|g_{\lambda}|\geq 2(q-1)q$. Denote by $N:=\#\{i\geq 4 :|g_{\lambda_i}|=q^2\}$. Since for $i\geq 4$, $|f_{\lambda_i}|=|g_{\lambda_i}|$ and $(q-3)2(q-1)q\geq(q-1)^3$, $N\geq1$. Furthermore, since $(q-3)q^2<(q-2)(q-1)q$, $N\leq q-4$. 

Assume that $|f_{\lambda_{N+4}}|\geq(N+1)q^2$. Then $$Nq^2+(q-3-N)(N+1)q^2\leq |f|<(q-1)^3$$ which gives $$Nq(q-N-3)< 3q-1.$$ This gives a contradiction since $1\leq N\leq q-4$. Furthermore the only possibility such that $|f_{\lambda_{N+4}}|=Nq^2$ is $N=1$ which is absurd since $f_{\lambda_{N+4}}$ is not a minimal weight codeword.

By Lemma \ref{2.15.1}, for all $4\leq i\leq N+3$, $g_{\lambda_{4}}=g_{\lambda_i}$. So, we can write for all $x=(x_1,\ldots,x_m)\in\F_q^m$ \begin{align*}f(x)&=\prod_{1\leq i\leq 3}(x_1-\lambda_i)\left(g_{\lambda_{4}}(x_2,\ldots,x_m)+ \prod_{4\leq i\leq N+3}(x_1-\lambda_i)h(x)\right)\\&=\prod_{1\leq i\leq 3}(x_1-\lambda_i)\left(\alpha f_{\lambda_{4}}(x_2,\ldots,x_m)+ \prod_{4\leq i\leq N+3}(x_1-\lambda_i)h(x)\right)\end{align*} where $h\in R_q((m-3)(q-1)-N,m)$ and $\alpha\in\F_q^*$. 

Then, for all $(x_2,\ldots,x_m)\in\F_q^{m-1}$, $$f_{\lambda_{N+4}}(x_2,\ldots,x_m)=\beta f_{\lambda_{4}}(x_2,\ldots,x_m)+\gamma h_{\lambda_{N+4}}(x_2,\ldots,x_m).$$ We get a contradiction by Lemma \ref{2.14}.

Now, assume $n=1$ or $n=2$.

If $(1-x_2^{q-1})$ divides $f$, we can write for all $x=(x_1,\ldots,x_m)\in\F_q^m$, $$f(x)=(1-x_2^{q-1})h(x_1,x_3,\ldots,x_m)$$ where $h\in R_q((m-4)(q-1)+3,m-1)$ and $|f|=|h|$. So, $h$ fulfils the same conditions as $f$. Iterating this process, we end either in the case where $t=0$ which is absurd or in the case where $(1-x_2^{q-1})$ does not divide $h$. So we can assume $(1-x_2^{q-1})$ does not divide $f$. 

Since $n\geq 1$, $f_{\lambda_1}=0$. So, $1-x_2^{q-1}$ divides $f_{\lambda_1}$. Then, since $1-x_2^{q-1}$ does not divide $f$, there exists $k\in\{1,\ldots,q-1\}$ such that for all $i\leq k$, $1-x_2^{q-1}$ divides $f_{\lambda_i}$ and $(1-x_2^{q-1})$ does not divide $f_{\lambda_{k+1}}$. For $\lambda\in\F_q$, if $|f_{\lambda}|>(q-3+n)q$ then $$|f_{\lambda}|\geq w_2=\left\{\begin{array}{ll}q^2&\textrm{if $n=2$}\\(q-1)^2&\textrm{if $n=1$}\end{array}\right..$$ Denote by $N:=\#\{i\geq n+1 :|f_{\lambda_i}|=(q-3+n)q\}$. In all cases, $(q-n)w_2\geq(q-1)^3.$ So, $N\geq1$.

Then, $|f_{\lambda_{n+1}}|=(q-3+n)q$ and $f_{\lambda_{n+1}}$ is a minimal weight codeword of $R_q((m-3)(q-1)+3-n,m-1)$ so, by applying an affine transformation, we can assume $1-x_2^{q-1}$ divides $f_{\lambda_{n+1}}$. Thus, $k\geq n+1\geq 2$.

If $1\leq k\leq n+N-1$, then $|f_{\lambda_{k+1}}|=(q-3+n)q<(q-3+k)q$. Otherwise, assume $|f_{\lambda_{k+1}}|\geq (q-3+k)q$. Then, $$|f|\geq N(q-3+n)q+(k-n-N)w_2+(q-k)(q-3+k)q.$$ In both cases, we get a contradiction since $|f|<(q-1)^3$ and $2\leq n+N\leq k\leq q-1$.

Since for all $n\leq i\leq k$, $1-x_2^{q-1}$ divides $f_{\lambda_i}$, it divides $g_{\lambda_i}$ too. Then we can write for all $x=(x_1,x_2,\ldots,x_m)\in\F_q^{m}$ \begin{align*}f(x)&=\prod_{1\leq i\leq n}(x_1-\lambda_i)(\prod_{n+1\leq i\leq k}(x_1-\lambda_i)h(x_1,x_2,x_3,\ldots,x_m)\\&\hspace{2cm}+(1-x_2^{q-1})l(x_1,x_3,\ldots,x_m))\end{align*} with $\deg(h)\leq (m-3)(q-1)+3-k$. Then for all $(x_2,\ldots,x_m)\in\F_q^{m-1}$, $$f_{\lambda_{k+1}}(x_2, \ldots,x_m)=\alpha h_{\lambda_{k+1}}(x_2,\ldots,x_m)+\beta(1-x_2^{q-1})l_{\lambda_{k+1}}(x_3,\ldots,x_m).$$ 
 Thus, we get a contradiction by Lemma \ref{3.5} since $k\geq2$ and $|f_{\lambda_{k+1}}|<(q-3+k)$.\end{preuve}
 
\subsection{General case}

\begin{proposition}\label{gene}Let $m\geq 2$, $q\geq 5$, $0\leq a\leq m-2$, $2\leq b\leq q-2$ and $f\in R_q(a(q-1)+b,m)$. Assume $c_b<(q-b+1)q$ and $b\neq 3$. If $|f|>(q-b+1)(q-1)q^{m-a-2}$ then $|f|\geq c_bq^{m-a-2}$.\end{proposition}

\begin{proposition}\label{gene3}Let $m\geq 3$, $q\geq7$, $0\leq a\leq m-3$. If $f\in R_q(a(q-1)+3,m)$ and $|f|>(q-1)(q-2)q^{m-a-2}$ then $|f|\geq (q-1)^3q^{m-a-3}$.\end{proposition}

We prove the two previous propositions in the same time. In order to simplify the notations, we set $$\widetilde{c}_b=\left\{\begin{array}{ll}c_b&\textrm{if $b\neq3$}\\(q-1)^3&\textrm{if $b=3$}\end{array}\right.$$

and $$m_0=\left\{\begin{array}{ll}2&\textrm{if $b\neq3$}\\3&\textrm{if $b=3$}\end{array}\right..$$

\begin{preuve} Lemmas \ref{t=0} and \ref{t=03} give the case where $a=0$. If $m=m_0$ we have considered all cases. Assume $m\geq m_0+1$ and $a\geq1$. We proceed by recursion on $a$. The case where $a=m-m_0$ comes from Lemmas \ref{m-2} and \ref{m-3}. If $m=m_0+1$, we have considered all cases. So, from now we assume $m\geq m_0+2$.
\\

Let $m-m_0-1\geq a\geq1$. Assume if $f\in R_q((a+1)(q-1)+b,m)$ is such that $|f|>(q-1)(q-b+1)q^{m-a-3}$ then, $|f|\geq \widetilde{c}_b q^{m-m_0-a-1}$. 

Let $f\in R_q(a(q-1)+b,m)$ such that $|f|>(q-b+1)(q-1)q^{m-a-2}$. Assume $|f|<\widetilde{c}_b q^{m-a-m_0}$ and denote by $S$ the support of $f$. 

Assume  $S$ meets all affine hyperplanes. We set an order on the elements of $\F_q$ such that $|f_{\lambda_1}|\leq\ldots\leq|f_{\lambda_q}|$. Since $|f|<\widetilde{c}_b q^{m-a-m_0}$, by recursion hypothesis, $f_{\lambda_1}$ is either a minimal weight codeword or a second weight codeword of $R_q(a(q-1)+b,m-1)$. In all cases, by applying an affine transformation, we can assume $(1-x_2^{q-1})$ divides $f_{\lambda_1}$. Let $1\leq k$ be such that for all $i\leq k$, $(1-x_2^{q-1})$ divides $f_{\lambda_i}$ but $(1-x_2^{q-1})$ does not divide $f_{\lambda_{k+1}}$. Then, by Lemma \ref{3.7}, $$|f|\geq (q-b)q^{m-a-1}+k(q-k)q^{m-a-2}\geq(q-b)q^{m-a-1}+(q-1)q^{m-a-2}.$$ We get a contradiction since $(q-b)q^{m-a-1}+(q-1)q^{m-a-2}\geq\widetilde{c}_b q^{m-a-m_0}$. 

So there exists an hyperplane $H_0$ which does not meet $S$. By applying an affine transformation we can assume $x_1=\alpha$, $\alpha\in\F_q$, is an equation of $H_0$. We denote by $n$ the number of hyperplanes parallel to $H_0$ which do not meet $S$. We set an order on the elements of $\F_q$ such that $|f_{\lambda_1}|\leq\ldots\leq |f_{\lambda_q}|$.

If $n=q-1$ then we can write for all $x=(x_1,\ldots,x_m)\in\F_q^m$, $$f(x)=(1-x_1^{q-1})g(x_2,\ldots,x_m)$$ where $g\in R_q((a-1)(q-1)+b,m-1)$ and $|f|=|g|$. So, $g$ fulfils the same conditions as $f$. Iterating this process, we end either in the case where $a=0$ (which gives a contradiction by Lemmas \ref{t=0} and \ref{t=03}) or in the case where $n<q-1$.

From now we assume $n<(q-1)$. By Lemma \ref{2.7}, since for $b\geq4$, $c_b\leq(q-2)(q-b+2)q^{m_0-2}$, for $b=2$, $(q^2-q-1)\leq 2(q-3)q$ and for $b=3$, $(q-1)^3\leq 2(q-4)q^2$, the only possibilities for $b\geq 2$ are $n=1$, $n=b-1$ or $n=b$.
 We can write for all $x=(x_1,\ldots,x_m)\in\F_q^m$ $$f(x)=\prod_{1\leq i\leq n}(x_1-\lambda_i)g(x)$$ where $g\in R_q(a(q-1)+b-n,m)$. Then for all $i\geq n+1$, $f_{\lambda_i}\in R_q(a(q-1)+b-n,m)$  and $|f_{\lambda_i}|=|g_{\lambda_i}|\geq (q-b+n)q^{m-a-2}$.

Assume $n=b$. For $\lambda\in\F_q$, if $|g_{\lambda}|>q^{m-a-1}$, then $|g_{\lambda}|\geq 2(q-1)q^{m-a-2}$. Denote by $N:=\#\{i\geq b+1 :|g_{\lambda_i}|=q^{m-a-1}\}$. Since for $i\geq b+1$, $|f_{\lambda_i}|=|g_{\lambda_i}|$ and $(q-b)2(q-1)q^{m-a-2}\geq \widetilde{c}_b q^{m-a-m_0}$, $N\geq1$. Furthermore, since $(q-b)q^{m-a-1}<(q-b+1)(q-1)q^{m-a-1}<|f|$, $N\leq q-b-1$. 

Assume $|f_{\lambda_{b+N+1}}|\geq(N+1)q^{m-a-1}$. Then $$Nq^{m-a-1}+(q-b-N)(N+1)q^{m-a-1}\leq |f|<\widetilde{c}_b q^{m-a-m_0}$$ which gives $$Nq^{m_0-1}(q-N-b)< \widetilde{c}_b -(q-b)q^{m_0-1}<q^{m_0-1}.$$ This gives a contradiction since $1\leq N\leq q-b-1$. Furthermore, the only possibility such that $|f_{\lambda_{b+N+1}}|=Nq^{m-a-1}$ is $N=1$ which is absurd since $f_{\lambda_{b+N+1}}$ is not a minimal weight codeword.

By Lemma \ref{2.15.1}, for all $b+1\leq i\leq N+b$, $g_{\lambda_{b+1}}=g_{\lambda_i}$. So, we can write for all $x=(x_1,\ldots,x_m)\in\F_q^m$ \begin{align*}f(x)&=\prod_{1\leq i\leq b}(x_1-\lambda_i)\left(g_{\lambda_{b+1}}(x_2,\ldots,x_m)+ \prod_{b+1\leq i\leq N+b}(x_1-\lambda_i)h(x)\right)\\&=\prod_{1\leq i\leq b}(x_1-\lambda_i)\left(\alpha f_{\lambda_{b+1}}(x_2,\ldots,x_m)+ \prod_{b+1\leq i\leq N+b}(x_1-\lambda_i)h(x)\right)\end{align*} where $h\in R_q(a(q-1)-N,m)$ and $\alpha\in\F_q^*$. 

Then, for all $(x_2,\ldots,x_m)\in\F_q^{m-1}$, $$f_{\lambda_{b+N+1}}(x_2,\ldots,x_m)=\beta f_{\lambda_{b+1}}(x_2,\ldots,x_m)+\gamma h_{\lambda_{b+N+1} }(x_2,\ldots,x_m).$$ We get a contradiction by Lemma \ref{2.14}.

Now, assume $n=1$ or $n=b-1$.

If $(1-x_2^{q-1})$ divides $f$, we can write for all $x=(x_1,\ldots,x_m)\in\F_q^m$, $$f(x)=(1-x_2^{q-1})g(x_1,x_3,\ldots,x_m)$$ where $g\in R_q((a-1)(q-1)+b,m-1)$ and $|f|=|g|$. So, $g$ fulfils the same conditions as $f$. Iterating this process, we end either in the case where $t=0$ which is impossible by Lemmas \ref{t=0} and \ref{t=03} or in the case where $(1-x_2^{q-1})$ does not divide $g$. So we can assume that $(1-x_2^{q-1})$ does not divides $f$. 

Since $n\geq 1$, $f_{\lambda_1}=0$. So, $1-x_2^{q-1}$ divides $f_{\lambda_1}$. Then, since $1-x_2^{q-1}$ does not divide $f$, there exists $k\in\{1,\ldots,q-1\}$ such that for all $i\leq k$, $1-x_2^{q-1}$ divides $f_{\lambda_i}$ and $(1-x_2^{q-1})$ does not divide $f_{\lambda_{k+1}}$. For $i\geq n+1$, if $|f_{\lambda_i}|>(q-b+n)q^{m-a-2}$ then $$|f_{\lambda_i}|\geq w_2=\left\{\begin{array}{ll}q^{m-a-1}&\textrm{if $n=b-1$}\\(q-b+2)(q-1)q^{m-a-3}&\textrm{if $n=1$}\end{array}\right..$$ Denote by $N:=\#\{i\geq n+1 :|f_{\lambda_i}|=(q-b+n)q^{m-a-2}\}$. If $4\leq b\leq \frac{q+3}{2}$, then $(q-n)w_2\geq(q-b+2)(q-2)q^{m-a-2}\geq c_bq^{m-a-2}$. If $b=2$ or $\frac{q}{2}+2\leq b\leq q-2$, $(q-n)w_2\geq((q-b+1)q-1)q^{m-a-2}$. If $b=3$, $(q-n)w_2\geq (q-1)^3q^{m-a-3}$. So, $N\geq1$. Since $(q-n)(q-b+n)q^{m-a-2}=(q-1)(q-b+1)q^{m-a-2}<|f|$, $N\leq q-n-1$. 

Then, $|f_{\lambda_{n+1}}|=(q-b+n)q^{m-a-2}$ and $f_{\lambda_{n+1}}$ is a minimal weight codeword of $R_q(a(q-1)+b-n,m-1)$ so, by applying an affine transformation, we can assume $1-x_2^{q-1}$ divides $f_{\lambda_{n+1}}$. Thus, $k\geq n+1\geq 2$.

If $1\leq k\leq n+N-1$, then $|f_{\lambda_{k+1}}|=(q-b+n)q^{m-a-2}<(q-b+k)q^{m-a-2}$. If $n+N\leq k\leq q-1$, assume $|f_{\lambda_{k+1}}|\geq (q-b+k)q^{m-a-2}$ Then $$|f|\geq N(q-b+n)q^{m-a-2}+(k-n-N)w_2+(q-k)(q-b+k)q^{m-a-2}$$ which gives a contradiction since $|f|<\widetilde{c}_b q^{m-a-m_0}$, $c_b<(q-b+1)q$ and $1\leq N\leq q-n-1$.

Since for all $n\leq i\leq k$, $1-x_2^{q-1}$ divides $f_{\lambda_i}$, it divides $g_{\lambda_i}$ too. Then we can write for all $x=(x_1,x_2,\ldots,x_m)\in\F_q^{m}$ \begin{align*}f(x)&=\prod_{1\leq i\leq n}(x_1-\lambda_i)(\prod_{n+1\leq i\leq k}(x_1-\lambda_i)h(x_1,x_2,x_3,\ldots,x_m)\\&\hspace{2cm}+(1-x_2^{q-1})l(x_1,x_3,\ldots,x_m))\end{align*} with $\deg(h)\leq a(q-1)+b-k$ and $l\in R_q((a-1)(q-1)+b-n,m-1)$. Then for all $(x_2,\ldots,x_m)\in\F_q^{m-1}$, $$f_{\lambda_{k+1}}(x_2, \ldots,x_m)=\alpha h_{\lambda_{k+1}}(x_2,\ldots,x_m)+\beta(1-x_2^{q-1})l_{\lambda_{k+1}}(x_3,\ldots,x_m).$$ 
 Thus, we get a contradiction by Lemma \ref{3.5} since $k\geq2$ and $|f_{\lambda_{k+1}}|<(q-b+k)q^{m-a-2}$.
\end{preuve}

\begin{theoreme}\label{w3}Let $m\geq 2$, $q\geq 5$, $0\leq a\leq m-2$, $2\leq b\leq q-2$. If $c_b<(q-b+1)q$ and either $b\neq3$ or $a\in\{0,m-2\}$ and $b=3$ then the third weight of $R_q(a(q-1)+b,m)$ is $W_3=c_bq^{m-a-2}$.  \end{theoreme}

\begin{theoreme}\label{w33}Let $m\geq 3$, $q\geq 5$, $0\leq a\leq m-3$. The third weight of $R_q(a(q-1)+b,m)$ is $W_3=(q-1)^3q^{m-a-3}$.\end{theoreme}

\begin{preuve}For $m=m_0$, it is the definition of ${c}_b$ and Lemma \ref{m=3}. Let $g\in R_q(b,m_0)$ such that $|g|=\widetilde{c}_b$. 

If $m\geq m_0+1$, by Proposition \ref{gene} and \ref{gene3} and Lemma \ref{m-2}, we have $W_3\geq \widetilde{c}_bq^{m-a-m_0}$.

For $x=(x_1,\ldots,x_m)\in\F_q^m$, we define $$f(x)=\prod_{i=1}^a(1-x_i^{q-1})g(x_{a+1},\ldots,x_{a+m_0}).$$ Then $f\in R_q(a(q-1)+b,m)$ and $|f|=|g|q^{m-a-2}=\widetilde{c}_bq^{m-2}$ which proves both theorems.
\end{preuve}

\section{The case where $m=2$}\label{cb}

\begin{theoreme}For $q\geq16$ and  $6\leq b<\frac{q+4}{3}$, $c_b=(q-b+2)(q-2)$.

Furthermore, if $f\in R_q(b,2)$ is such that $|f|=(q-b+2)(q-2)$ then up to affine transformation for all $(x,y)\in\F_q^2$,  $$f(x,y)=\prod_{i=1}^{b-2}(x-b_i)(y-c)(y-d)$$ where $b_i\in \F_q$ are such that for $i\neq j$, $b_i\neq b_j$, $c\in\F_q$, $d\in\F_q$ and $c\neq d$,

or 

$$ f(x,y)=\prod_{i=1}^{b-1}(a_i x+b_i y)(a_1x+b_1y+e)$$ where $(a_i,b_j)\in\F_q^2\setminus\{(0,0)\}$, for $i\neq j$, $a_ib_j-a_jb_i\neq0$ and $e\in\F_q^*$

or

$$f(x,y)=\prod_{i=1}^{3}(a_i x+b_i y)\prod_{j=1}^{b-3}(a_1x+b_1y+e_j)$$ where $(a_i,b_j)\in\F_q^2\setminus\{(0,0)\}$, for $i\neq j$, $a_ib_j-a_jb_i\neq0$, $e_i\in\F_q^*$ and $e_i\neq e_j$ for $j\neq i$.

\end{theoreme}

\begin{preuve}Let $6\leq b<\frac{q+4}{3}$ and $f\in R_q(b,2)$ such that $|f|=c_b$ and denote by $S$ its support.

From section \ref{upper}, we know that in this case $c_b\leq (q-b+2)(q-2)<(q-b+1)q$. Since $(q-b+1)q$ is the minimum weight of $R_q(b-1,2)$, $\deg(f)=b$. We prove first that $f$ is the product of $b$ affine factors. 
Let $P$ be a point of $\F_q^2$ which is not in $S$ and $L$ be a line in $\F_q^2$ such that $P\in L$. Then, either $L$ does not meet $S$ or $L$ meets $S$ in at least $q-b$ points. If any line through $P$ meets $S$ then $$(q+1)(q-b)\leq|f|\leq(q-b+2)(q-2)$$ which is absurd since $b< \frac{q+4}{3}$. So there exists a line through $P$ which does not meet $S$. By applying the same argument to all $P$ not in $S$, we get that $f$ is the product of affine factors.

We have just proved that $Z$ the set of zeros of $f$ is the union of $b$ lines in $\F_q^2$. We say that those lines are in configuration $A_b$ is the $b$ lines are parallel, in configuration $B_b$ if exactly $b-1$ lines are parallel, in configuration $C_b$ if the $b$ lines meet in a point, in configuration $D_b$ if $b-2$ lines are parallel and the 2 other lines are also parallel, in configuration $E_b$ if $b-2$ lines are parallel and the 2 other lines intersect in one point included in one of the parallel lines, in configuration $F_b$ if $b-1$ lines intersect in one point and the $b$th line is parallel to one of the previous. We say that we are in configuration $G_b$ if we are in none of the previous configurations.

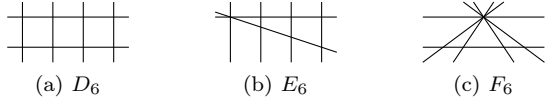
\begin{figure}[h!]
\begin{center}\subfloat[$D_6$]{\begin{tikzpicture}[scale=0.2]
\draw (0,3)--(8,3);
\draw (0,1)--(8,1);
\draw (1,4)--(1,0);
\draw (3,4)--(3,0);
\draw (5,0)--(5,4);
\draw (7,0)--(7,4);
\end{tikzpicture}}
\hspace{1cm}
\subfloat[$E_6$]{\begin{tikzpicture}[scale=0.2]
\draw (0,3)--(8,3);
\draw (0,20/6)--(8,4/6);
\draw (1,4)--(1,0);
\draw (3,4)--(3,0);
\draw (5,0)--(5,4);
\draw (7,0)--(7,4);
\end{tikzpicture}}
\hspace{1cm}
\subfloat[$F_6$]{\begin{tikzpicture}[scale=0.2]
\draw (0,3)--(8,3);
\draw (0,1)--(8,1);
\draw (0,0)--(16/3,4);
\draw (2,0)--(14/3,4);
\draw (6,0)--(10/3,4);
\draw (8,0)--(8/3,4);
\end{tikzpicture}}
\end{center}\end{figure}
We prove by induction on $b$ that $Z$ the set of zeros of $f$ is of type $D_b$, $E_b$ or $F_b$.
Since the cardinal of such set is $(q-b+2)(q-2)$, by Lemma \ref{DGMW1} we get the result.

For $b=6$, denote by $Z$ the set of the zeros of $f$. We have just proved that $Z$ is the union of 6 lines in $\F_q^2$. If the 6 lines are parallel then $f$ is minimum weight codeword of $R_q(6,2)$ which is absurd. If 5 of these lines are parallel or the 6 lines intersect in a point then, $f$ is a second weight codeword of $R_q(6,2)$ which is absurd. 
If 4 of these lines are parallel, then if the 2 other lines are parallel and we are in configuration $D_6$, if 3 of theses lines intersect in a point then we are in configuration $E_6$ otherwise $\#Z=6q-9<q^2-(q-4)(q-2)=6q-8$ which is absurd since $|f|\leq(q-4)(q-2)$. If 3 of theses lines are parallel then, they intersect the 3 other lines so $\#Z\leq 6q-9<6q-8$ which is absurd. 
If 2 of these lines are parallel then, if at least two of the other lines intersect in a point which is not included in one of the parallel lines then $\#Z\leq6q-9$ which is absurd. So the only possibility in this case is configuration $F_6$. If all lines intersect pairwise then they cannot intersect in one point, so $\#Z\leq 6q-9$ which is absurd. This proves the result for $b=6$.  

Let $6\leq b<\frac{q+1}{2}$. Assume if $f\in R_q(b,2)$ and $|f|=c_b$ then its set of zeros is of type $D_b$, $E_b$ or $F_b$.

Let $f\in R_q(b+1,2)$ such that $|f|=c_b\leq(q-b+1)(q-2)=q^2-(bq+q-2b+2)$. Denote by $Z$ the set of zeros of $f$, it is the union of $b+1$ lines in $\F_q^2$. Suppose that $Z$ is of type $G_{b+1}$. We decompose $Z$ in a configuration of $b$ lines and a line. We have 6 possible cases : 
\begin{itemize}
\item $Z$ is the union of a type $G_b$ configuration and a line. Since a configuration $G_b$ is composed of neither at least $b-1$ parallel lines nor $b$ lines which meet in the same point. The line meets the configuration $G_b$ in at least 2 points. Then, by induction hypothesis, $\#Z<bq-2b+4+q-2=bq+q-2b+2$.
\item $Z$ is the union of a type $F_b$ configuration and a line. Since $Z$ is a configuration $G_{b+1}$, the line cannot intersect the configuration $F_b$ in the point where $b-1$ lines of the configuration intersect. So, the line intersects the configuration $F_b$ in at least 3 points and $\#Z\leq bq-2b+4+q-3=bq+q-2b+1$.
\item $Z$ is the union of a type $E_b$ configuration and a line. Since $Z$ is a configuration $G_{b+1}$, the line cannot be parallel to the $b-2$ lines parallel in the configuration $E_b$. So, the line intersects the configuration $E_b$ in at least 3 points and $\#Z\leq bq-2b+4+q-3=bq+q-2b+1$.
\item $Z$ is the union of a type $D_b$ configuration and a line. Since $Z$ is a configuration $G_{b+1}$, the line cannot be parallel to the $b-2$ lines parallel in the configuration $D_b$. So, the line intersects the configuration $D_b$ in at least 3 points and $\#Z\leq bq-2b+4+q-3=bq+q-2b+1$.
\item $Z$ is the union of a type $C_b$ configuration and a line. Since $Z$ is a configuration $G_{b+1}$, the line can neither be parallel to one of the lines in the configuration $C_b$ nor intersect $C_b$ in the point where all the lines of $C_b$ intersect. So, the line intersects the configuration $C_b$ in at least $b$ points and $\#Z\leq bq-b+1+q-b=bq+q-2b+1$.
\item $Z$ is the union of a type $B_b$ configuration and a line. Since $Z$ is a configuration $G_{b+1}$, the line can neither be parallel to one of the lines in the configuration $B_b$ nor intersect the configuration $B_b$ in a point included in 2 different lines. So, the line intersects the configuration $B_b$ in at least $b$ points and $\#Z\leq bq-b+1+q-b=bq+q-2b+1$.
\end{itemize}  
Since $A_{b+1}$, $B_{b+1}$ and $C_{b+1}$ are minimal or second weight configurations, the zeros of $f$ are of type $D_{b+1}$, $E_{b+1}$ or $F_{b+1}$.\end{preuve}

\begin{proposition}For $q\geq9$, $c_4=(q-2)^2$.

Furthermore, if $f\in R_q(4,2)$ is such that $|f|=(q-2)^2$ then up to affine transformation for all $(x,y)\in\F_q^2$, either $$f(x,y)=(x-a)(x-b)(y-c)(y-d)$$ where $a\in\F_q$, $b\in\F_q$, $c\in\F_q$, $d\in\F_q$ are such that $a\neq b$ and  $c\neq d$

or 

$$ f(x,y)=\prod_{i=1}^{3}(a_i x+b_i y)(a_1x+b_1y+e)$$ where $(a_i,b_j)\in\F_q^2\setminus\{(0,0)\}$, for $i\neq j$, $a_ib_j-a_jb_i\neq0$ and $e\in\F_q^*$.\end{proposition}

\begin{preuve}Let $f\in R_q(4,2)$ such that $|f|=c_4$ and denote by $S$ its support.

From section \ref{upper}, we know that in this case $c_4\leq (q-2)^2<(q-3)q$. Since $(q-3)q$ is the minimum weight of $R_q(3,2)$, $\deg(f)=4$. We prove first that $f$ is the product of $4$ affine factors. 
Let $P$ be a point of $\F_q^2$ which is not in $S$ and $L$ be a line in $\F_q^2$ such that $P\in L$. Then, either $L$ does not meet $S$ or $L$ meets $S$ in at least $q-4$ points. If any line through $P$ meets $S$ then $$(q+1)(q-4)\leq|f|\leq(q-2)^2$$ which is absurd since $q\geq 9$. So there exists a line through $P$ which does not meet $S$. By applying the same argument to all $P$ not in $S$, we get that $f$ is the product of affine factors.

Denote by $Z$ the set of zeros of $f$. We have just proved that $Z$ is the union of 4 lines in $\F_q^2$. If the 4 lines are parallel then $f$ is minimum weight codeword of $R_q(4,2)$ which is absurd. If 3 of these lines are parallel or the 4 lines intersect in a point then, $f$ is a second weight codeword of $R_q(4,2)$ which is absurd. 
Assume 2 of these lines are parallel. If the 2 other lines are parallel and we are in the first case of the proposition. If 3 of the 4 lines intersect in a point then we are in the second case of the proposition, otherwise $\#Z=4q-5<q^2-(q-2)^2=4q-4$ which is absurd since $|f|\leq(q-2)^2$.
Assume all lines intersect pairwise. They cannot intersect in one point, so $\#Z\leq 4q-5$ which is absurd. \end{preuve}

\begin{proposition}For $q\geq13$, $c_5=(q-3)(q-2)$.

Furthermore, if $f\in R_q(5,2)$ is such that $|f|=(q-3)(q-2)$ then up to affine transformation for all $(x,y)\in\F_q^2$,  $$f(x,y)=\prod_{i=1}^{3}(x-b_i)(y-c)(y-d)$$ where $b_i\in \F_q$ are such that for $i\neq j$, $b_i\neq b_j$, $c\in\F_q$, $d\in\F_q$ and $c\neq d$

or 

$$ f(x,y)=\prod_{i=1}^{4}(a_i x+b_i y)(a_1x+b_1y+e)$$ where $(a_i,b_j)\in\F_q^2\setminus\{(0,0)\}$, for $i\neq j$, $a_ib_j-a_jb_i\neq0$ and $e\in\F_q^*$

or

$$f(x,y)=\prod_{i=1}^{3}(a_i x+b_i y)\prod_{j=1}^{2}(a_1x+b_1y+e_j)$$ where $(a_i,b_j)\in\F_q^2\setminus\{(0,0)\}$, for $i\neq j$, $a_ib_j-a_jb_i\neq0$, $e_i\in\F_q^*$ and $e_i\neq e_j$ for $j\neq i$

or 

$$f(x,y)=(x-a)(x-b)(y-c)(y-d)\left((d-c)x+(a-b)y+bc-ad\right)$$ where $a\in\F_q$, $b\in\F_q$, $c\in\F_q$, $d\in\F_q$ are such that $a\neq b$ and  $c\neq d$\end{proposition}

\begin{preuve}Let $f\in R_q(5,2)$ such that $|f|=c_5$ and denote by $S$ its support.

From section \ref{upper}, we know that in this case $c_5\leq (q-3)(q-2)<(q-4)q$. Since $(q-4)q$ is the minimum weight of $R_q(4,2)$, $\deg(f)=5$. We prove first that $f$ is the product of $5$ affine factors. 
Let $P$ be a point of $\F_q^2$ which is not in $S$ and $L$ be a line in $\F_q^2$ such that $P\in L$. Then, either $L$ does not meet $S$ or $L$ meets $S$ in at least $q-5$ points. If any line through $P$ meets $S$ then $$(q+1)(q-5)\leq|f|\leq(q-3)(q-2)$$ which is absurd since $q\geq13$. So there exists a line through $P$ which does not meet $S$. By applying the same argument to all $P$ not in $S$, we get that $f$ is the product of affine factors.

Denote by $Z$ the set of zeros of $f$. We have just proved that $Z$ is the union of 5 lines in $\F_q^2$. If the 5 lines are parallel then $f$ is minimum weight codeword of $R_q(5,2)$ which is absurd. If 4 of these lines are parallel or the 5 lines intersect in one point then, $f$ is a second weight codeword of $R_q(5,2)$ which is absurd. 
Assume 3 of these lines are parallel. If the 2 other lines are parallel and we are in the first case of the proposition. If 3 of the 5 lines lines intersect in a point then we are in the second case of the proposition, otherwise $\#Z=5q-7<q^2-(q-3)(q-2)=5q-6$ which is absurd since $|f|\leq(q-3)(q-2)$. 
Assume 2 of the lines are parallel. If an other pair of lines is parallel, then either we are in the last case of the proposition or the fifth line meets the four other lines in at least 3 points and $\#Z\leq 4q-4+q-3=5q-7$ which is absurd. If 4 of the 5 lines intersect in a point then we are in the third case of the proposition, otherwise $\#Z=5q-7<5q-6$ which is absurd.
Assume all lines intersect pairwise. They cannot intersect in one point, so $\#Z\leq 5q-7$ which is absurd.\end{preuve}

\section{Codeword reaching the third weight}

\begin{proposition}\label{case1}Let $q\geq 5$, $m\geq 2$, $4\leq b\leq q-2$, $f\in R_q(b,m)$ and $g\in R_q(b,2)$ such that $|g|=c_b$. If $c_b<(q-b+1)q-1$ and $|f|=c_bq^{m-2}$ then up to affine transformation, for all $x=(x_1,\ldots,x_m)\in\F_q^m$, $f(x)=g(x_1,x_2)$. \end{proposition}

\begin{preuve}The case where $m=2$ comes from the definition of $c_b$. Assume $m\geq 3$. 

Let $f\in R_q(b,m)$ such that $4\leq b<q-1$ and $|f|=c_bq^{m-2}$. Assume $c_b<(q-b+1)q-1$. We denote by $S$ the support of $f$.

Suppose first that $f$ has an affine factor. By applying an affine transformation, we can assume that $x_1$ divides $f$. Let $n=\#\{\lambda\in\F_q:(x_1-\lambda)\textrm{ divides }f\}$. Since $\deg(f)\leq b$ and $f$ is neither a minimal weight codeword nor a second weight codeword, $n\leq b-2$ (see proof of Lemma \ref{t=0}). Furthermore, if $b\geq \frac{q+3}{2}$, $(q-b+1)-1\leq (q-b+2)(q-2)$ then, by Lemma \ref{2.7}, the only possibility for $n$ is 1. If $b<\frac{q+3}{2}$, $c_b\leq (q-b+2)(q-2)$. So, by Lemma \ref{2.7}, the only possibility for $n$ are 1, 2, $b-2$.

We can write for all $x=(x_1,\ldots,x_m)\in \F_q$, $$f(x)=\prod_{1=1}^n(x_1-\lambda_i)g(x)$$ with $\lambda_i\in\F_q$, $\lambda_i\neq\lambda_j$ for $i\neq j$ and $\deg(g)\leq b-n$. Then for $i\geq n+1$, $\deg(f_{\lambda_i}) \leq b-n$ and $|f_{\lambda_i}|\geq(q-b+n)q^{m-2}$.

If $n=b-2$ or $n=2$ Then, for reasons of cardinality, for all $i\geq n+1$, $|f_{\lambda_i}|=(q-b+n)q^{m-2}$ and $f_{\lambda_i}$ is a minimum weight codeword of $R_q(b-n,m-1)$.

If $n=1$, we denote by $N_1:=\#\{i\geq n+1:|f_{\lambda_i}|=(q-b+1)q^{m-2}\}$. For $i\geq n+1$, if $|f_{\lambda_i}|>(q-b+1)q^{m-2}$ then, $|f_{\lambda_i}|\geq (q-b+2)(q-1)q^{m-3}$. If $b<\frac{q+3}{2}$, $(q-1)^2(q-b+2)q^{m-3}>(q-b+2)(q-2)q^{m-2}$. If $b\geq\frac{q+3}{2}$, $(q-1)^2(q-b+2)q^{m-3}\geq ((q-b+1)q-1)q^{m-2}$. So, in both cases, $N_1\geq1$. 

Suppose now that $f$ meets all hyperplanes ($n=0$). Since $(q-b+1)(q-1)<c_b<(q-b+1)q$, $c_b\not\equiv 0\mod q$. Then by Lemma \ref{inter}, there exists $H$ an hyperplane such that either $\#(S\cap H)=(q-b)q^{m-2}$ or $\#(S\cap H)=(q-b+1)(q-1)q^{m-3}$. If $\#(S\cap H)=(q-b+1)(q-1)q^{m-3}$ then there exists $A$ an affine subspace of codimension 2 included in $H$ which does not meet $S$. If all hyperplanes through $A$ meet $S$ in $(q-b+1)(q-1)q^{m-3}$ points then, $|f|\geq(q+1)(q-1)(q-b+1)q^{m-3}\geq((q-b+1)q-1)q^{m-2}$ which gives a contradiction. So there exists an hyperplane which meets $S$ in $(q-b)q^{m-2}$ points. 

In all cases, there is an hyperplane which meets $S$ in $(q-b+n)q^{m-2}$ points. By applying an affine transformation, we can assume $x_1=\lambda$, $ \lambda\in\F_q$ is an equation of this hyperplane.

We set an order on the elements of $\F_q$ such that $|f_{\lambda_1}|\leq\ldots\leq|f_{\lambda_q}|$, then $f_{\lambda_{n+1}}$ is a minimum weight codeword of $R_q(b_n,m-1)$. By applying an affine transformation, we can assume $f_{\lambda_{n+1}}$ depends only on $x_2$. Let $k\in\{n+1,\ldots,q\}$ be such that for all $i\leq k$, $f_{\lambda_i}$ depends only on $x_2$ and $f_{\lambda_{k+1}}$ does not depend only on $x_2$.

If $k>b$ then for all $x=(x_1,\ldots,x_m)\in\F_q^m)$, $$f(x)=\sum_{i=0}^bf_{\lambda_{i+1}}^{(i)}(x_2,\ldots,x_m)\prod_{1\leq j\leq i}(x_1-\lambda_j).$$
Since for $i \leq b+1$, $f_{\lambda_i}$ depends only on $x_2$ then $f$ depends only on $x_1$ and $x_2$ which proves the proposition in this case.

If $b\geq k$ Then we can write for all $x=(x_1,\ldots,x_m)\in\F_q^m)$, $$f(x)=g(x_1,x_2)+\prod_{i=1}^k(x_1-\lambda_i)h(x)$$ with $\deg(h)\leq b-k$.
Then, for all $(x_2,\ldots,x_m)\in\F_q^{m-1}$, $$f_{\lambda_{k+1}}(x_2,\ldots,x_m)=g_{\lambda_{k+1}}(x_2)+\alpha h_{\lambda_{k+1}}(x_2,\ldots,x_m)$$ with $\alpha\in\F_q^*$. Since $f_{\lambda_{k+1}}$ does not depend only on $x_2$, by Lemma \ref{3.9}, we have $|f_{\lambda_{k+1}}|\geq(q-b+k)q^{m-2}$. If $n=2$ or $n=b-2$, we get a contradiction since $k\geq n+1$ and $|f_{\lambda_{k+1}}|=(q-b+n)q^{m-2}$. If $n=0$ or $n=1$, we get $$|f|\geq (k-n)(q-b+n)q^{m-2}+(q-k)(q-b+k)q^{m-2}$$ which is absurd since for $n+1\leq k\leq b$, $$(k-n)(q-b+n)q^{m-2}+(q-k)(q-b+k)q^{m-2}\geq((q-b+1)q-1)q^{m-2}.$$
\end{preuve}

\begin{lemme}\label{33}Let $q\geq7$, $f\in R_q(3,3)$ such that $|f|=(q-1)^3$ then up to affine transformation, for all $x=(x_1,x_2,x_3)\in\F_q^3$, $$f(x)=x_1x_2x_3.$$\end{lemme}

\begin{preuve}Since $(q-1)^3>(q-2)q^2$, $\deg(f)=3$. We prove first that $f$ is the product of $3$ affine factors. 
Let $P$ be a point of $\F_q^3$ which is not in $S$ and $L$ be a line in $\F_q^3$ such that $P\in L$. Then, either $L$ does not meet $S$ or $L$ meets $S$ in at least $q-3$ points. If any line through $P$ meets $S$ then $$(q^2+q+1)(q-3)\leq|f|=(q-1)^3$$ which is absurd since $q\geq7$. So there exists $L_P$ a line through $P$ which does not meet $S$. Then, let $A$ be plane through $L_P$, either $A$ does not meet $S$ or $A$ meets $S$ in at least $(q-3)q$ points. So, $(q+1)(q-3)q\leq|f|=(q-1)^3$ which is absurd since $q\geq 7$. So there exists $A_P$ a plane containing $P$ which does not meet $S$. By applying the same argument to all $P$ not in $S$, we get that $f$ is the product of affine factors.

We have just prove that $Z$ the set of zeros of $f$ is the union of $3$ planes. If these 3 planes are parallel, $f$ is a minimum weight codeword. If 2 of these planes are parallel or the 3 planes intersect in a line, we get a second weight codeword. So the 3 planes intersect pairwise in a line. If the 3 intersection are parallel lines (see figure \ref{mot33}) then $\#Z=3q^2-3q$ which is absurd. The only possibility left gives the result.

\begin{figure}[h!]\caption{}\label{mot33}
\begin{center}\begin{tikzpicture}[scale=0.2]
\draw[dashed] (1,1)--(1,7);
\draw[dashed] (7,4)--(7,10);
\draw[dashed] (3,9)--(3,15);
\draw (0,39/6)--(8,63/6);
\draw (26/8,10)--(26/8,16);
\draw (26/8,16)--(6/8,6);
\draw (6/8,0)--(6/8,6);
\draw (6/8,0)--(26/8,10);
\draw (0,39/6-6)--(8,63/6-6);
\draw (0,39/6)--(0,39/6-6);
\draw (8,63/6-6)--(8,63/6);
\draw (11/5,16)--(11/5,10);
\draw (11/5,16)--(39/5,9);
\draw (39/5,9)--(39/5,3);
\draw (39/5,3)--(11/5,10);
\end{tikzpicture}
\end{center}
\end{figure}\end{preuve}

\begin{proposition}\label{case2}Let $q\geq 7$, $m\geq 3$, $f\in R_q(3,m)$. If $|f|=(q-1)^3q^{m-3}$ then up to affine transformation, for all $x=(x_1,\ldots,x_m)\in\F_q^m$, $f(x)=x_1x_2x_3$. \end{proposition}

\begin{preuve}The case where $m=3$ comes from Lemma \ref{33}. Assume $m\geq 4$. 

Let $f\in R_q(3,m)$ such that $|f|=(q-1)^3q^{m-3}$. We denote by $S$ the support of $f$.

Suppose first that $f$ has an affine factor. By applying an affine transformation, we can assume $x_1$ divides $f$. Let $n=\#\{\lambda\in\F_q:(x_1-\lambda)\textrm{ divides }f\}$. Since $\deg(f)\leq b$ and $f$ is neither a minimal weight codeword nor a second weight codeword, $n=1$.

We can write for all $x=(x_1,\ldots,x_m)\in \F_q$, $$f(x)=x_1g(x)$$ with $\deg(g)\leq 2$. Then for $\lambda\in\F_q^*$, $\deg(f_{\lambda}) \leq 2$ and $|f_{\lambda}|\geq(q-2)q^{m-2}$.

We denote by $N_1:=\#\{\lambda\in\F_q^*:|f_{\lambda}|=(q-2)q^{m-2}\}$. For $\lambda\in\F_q^*$, if $|f_{\lambda}|>(q-2)q^{m-2}$ then, $|f_{\lambda}|\geq (q-1)^2q^{m-3}$. So either $N_1\geq1$ or for all $\lambda\in\F_q^*$, $|f_{\lambda}|=(q-1)^2q^{m-3}$. 

If for all $\lambda\in\F_q^*$, $|f_{\lambda}|=(q-1)^2q^{m-3}$ then for any $\lambda\in\F_q^*$, there are two non parallel affine subspaces of codimension 2 included in the hyperplane of equation $x_1=\lambda$ which do not meet $S$. Let $A$ be one of these affine subspace in $x_1=\lambda'$. Since $2(q-2)>q+1$, there exists $\lambda''\in\F_q^*$ such that $A_1$ one of the affine subspace of codimension 2 included in $x_1=\lambda''$ is parallel to $A$. Then $H_0$ the hyperplane through $A$ and $A_1$ contains at least one more point which is in an hyperplane of equation $x_1=\lambda$, $\lambda\neq\lambda'$ and $\lambda\neq\lambda''$. So $H_0$ meets $S$ in at most $q^{m-1}-3q^{m-2}-1$. Since the minimum weight of $R_q(3,m-1)$ is $(q-3)q^{m-2}$, $H_0$ does not meet $S$. Applying the same argument to the other affine subspace of codimension 2 included in the hyperplane of equation $x_1=\lambda'$. We get that $f$ is the product of 3 linear factors. So, $Z$ the set of zeros of $f$ is the union of $3$ hyperplanes. If these 3 hyperplanes are parallel, $f$ is a minimum weight codeword. If 2 of these hyperplanes are parallel or those 3 hyperplanes intersect in an affine subspace of codimension 2, we get a second weight codeword. So the 3 planes intersect pairwise in an affine subspace of codimension 2. If the 3 intersections are parallel affine subspaces then $\#Z=3q^{m-1}-3q^{m-2}$ which is absurd. The only possibility left gives the result.

From now, we suppose that if $n=1$ then $N_1\geq1$.

Suppose now that $f$ meets all hyperplanes ($n=0$). Since $(q-1)^3\not\equiv 0\mod q$, by Lemma \ref{inter}, there exists $H$ an hyperplane such that either $\#(S\cap H)=(q-3)q^{m-2}$ or $\#(S\cap H)=(q-2)(q-1)q^{m-3}$. If $\#(S\cap H)=(q-2)(q-1)q^{m-3}$ then there exists $A$ an affine subspace of codimension 2 included in $H$ which does not meet $S$. If all hyperplanes through $A$ meet $S$ in $(q-2)(q-1)q^{m-3}$ points then, $|f|\geq(q+1)(q-1)(q-2)q^{m-3}>(q-1)^3q^{m-3}$ which gives a contradiction. So there exists an hyperplane which meets $S$ in $(q-3)q^{m-2}$ points.

In all cases, there exists an hyperplane which meets $S$ in $(q-b+N)q^{m-2}$ points. By applying an affine transformation, we can assume $x_1=\lambda$, $\lambda\in\F_q$ is an equation of this hyperplane. We set an order on the elements of $\F_q$ such that $|f_{\lambda_1}|\leq\ldots\leq|f_{\lambda_q}|$, then $f_{\lambda_{n+1}}$ is a minimum weight codeword of $R_q(3-n,m-1)$. By applying an affine transformation, we can assume $f_{\lambda_{n+1}}$ depends only on $x_2$. Let $k\in\{n+1,\ldots,q\}$ be such that for all $i\leq k$, $f_{\lambda_i}$ depends only on $x_2$ and $f_{\lambda_{k+1}}$ does not depend only on $x_2$.

If $k>b$ then for all $x=(x_1,\ldots,x_m)\in\F_q^m)$, $$f(x)=\sum_{i=0}^3f_{\lambda_{i+1}}^{(i)}(x_2,\ldots,x_m)\prod_{1\leq j\leq i}(x_1-\lambda_j).$$
Since for $i \leq 4$, $f_{\lambda_i}$ depends only on $x_2$ then $f$ depends only on $x_1$ and $x_2$ which gives a contradiction since $(q-1)^3q^{m-3}\not \equiv0\mod q^{m-2}$.

If $3\geq k$ Then we can write for all $x=(x_1,\ldots,x_m)\in\F_q^m)$, $$f(x)=g(x_1,x_2)+\prod_{i=1}^k(x_1-\lambda_i)h(x)$$ with $\deg(h)\leq 3-k$.
Then, for all $(x_2,\ldots,x_m)\in\F_q^{m-1}$, $$f_{\lambda_{k+1}}(x_2,\ldots,x_m)=g_{\lambda_{k+1}}(x_2)+\alpha h_{\lambda_{k+1}}(x_2,\ldots,x_m)$$ with $\alpha\in\F_q^*$. Since $f_{\lambda_{k+1}}$ does not depend only on $x_2$, by Lemma \ref{3.9}, we have $|f_{\lambda_{k+1}}|\geq(q-3+k)q^{m-2}$. Then, we get $$|f|\geq (k-n)(q-b+n)q^{m-2}+(q-k)(q-b+k)q^{m-2}$$ which is absurd since for $n+1\leq k\leq b$, $$(k-n)(q-b+n)q^{m-2}+(q-k)(q-b+k)q^{m-2}>(q-1)^3q^{m-3}.$$\end{preuve}

\begin{lemme}\label{suphyp}Let $m\geq3$, $q\geq7$, $1\leq a\leq m-2$ and $4\leq b \leq q-2$. If $f\in R_q(a(q-1)+b,m)$ is such that $|f|=c_bq^{m-a-2}$ and $c_b<(q-b+1)q-1$, then the support of $f$ is included in an affine hyperplane of $\F_q^m$.\end{lemme}

\begin{lemme}\label{suphyp2}Let $m\geq4$, $q\geq7$, $1\leq a\leq m-3$. If $f\in R_q(a(q-1)+3,m)$ is such that $|f|=(q-1)^3q^{m-a-3}$, then the support of $f$ is included in an affine hyperplane of $\F_q^m$.\end{lemme}

We set $m_0$ and $\widetilde{c}_b$ as in Section \ref{poids3}. We prove both lemmas in the same time.\\

\begin{preuve} Let $f\in R_q(a(q-1)+b,m)$ such that $|f|=\widetilde{c}_bq^{m-a-m_0}$ and denote by $S$ the support of $f$. Assume $S$ is not included in a hyperplane.

Assume $S$ meets all affine hyperplanes.  If $m=3$, since $(q-b+1)q>\widetilde{c}_b$ necessarily, there exists an hyperplane which meets $S$ in $(q-b)$ points. Assume $m\geq 4$, since $(q-b+1)(q-1)<{c}_b<(q-b+1)q$, $c_b\not\equiv0\mod q$. Since $(q-1)^3\not\equiv 0\mod q$, by Lemma \ref{inter}, in both cases there exists an hyperplane which meets $S$ in $(q-b)q^{m-a-2}$ points or $(q-b+1)(q-1)q^{m-a-3}$ points. By applying an affine transformation, we can assume $x_1=\lambda$, $\lambda\in\F_q$ is an equation of this hyperplane. We set an order on the elements of $\F_q$ such that $|f_{\lambda_1}|\leq\ldots\leq |f_{\lambda_q}|$. Then, by applying an affine transformation, we can assume that $(1-x_2^{q-1})$ divides $f_{\lambda_1}$. Let $1\leq k$ be such that for all $i\leq k$, $(1-x_2^{q-1})$ divides $f_{\lambda_i}$ but $(1-x_2^{q-1})$ does not divide $f_{\lambda_{k+1}}$. Then, by Lemma \ref{3.7}, $|f|\geq (q-b)q^{m-a-1}+k(q-k)q^{m-a-2}\geq(q-b)q^{m-a-1}+(q-1)q^{m-a-2}$. We get a contradiction since $c_b<(q-b+1)q-1$. 

So there exists an hyperplane $H_0$ which does not meet $S$. By applying an affine transformation we can assume $x_1=\alpha$, $\alpha\in\F_q$, is an equation of $H_0$. We denote by $n$ the number of hyperplanes parallel to $H_0$ which does not meet $S$. We set an order on the elements of $\F_q$ such that $|f_{\lambda_1}|\leq\ldots\leq |f_{\lambda_q}|$.

Since $S$ is not included in an hyperplane, $n<q-1$. If $b\geq\frac{q+3}{2}$ then $c_b<(q-b+2)(q-2)q^{m_0-2}$ and if $b=3$, $(q-1)^3<\leq2(q-4)q^2$. So, by Lemma \ref{2.7}, for $b=3$ or $b\geq \frac{q+3}{2}$, the only possibilities are $n=1$, $n=b-1$ or $n=b$. If $4\leq b<\frac{q+3}{2}$ then, $\widetilde{c}_b\leq(q-b+2)(q-2)$. So, by Lemma \ref{2.7}, the only possibilities are $n=1$, $n=2$, $n=b-2$, $n=b-1$ or $n=b$.
 We can write for all $x=(x_1,\ldots,x_m)\in\F_q^m$ $$f(x)=\prod_{1\leq i\leq n}(x_1-\lambda_i)g(x)$$ where $g\in R_q(a(q-1)+b-n,m)$. Then for all $i\geq n+1$, $f_{\lambda_i}\in R_q(a(q-1)+b-n,m)$  and $|f_{\lambda_i}|=|g_{\lambda_i}|\geq (q-b+n)q^{m-a-2}$.

Assume $n=b$. For $\lambda\in\F_q$, if $|g_{\lambda}|>q^{m-a-1}$, then $|g_{\lambda}|\geq 2(q-1)q^{m-a-2}$. Denote by $N:=\#\{i\geq b+1 :|g_{\lambda_i}|=q^{m-a-1}\}$. Since for $i\geq b+1$, $|f_{\lambda_i}|=|g_{\lambda_i}|$ and $(q-b)2(q-1)q^{m-a-2}> \widetilde{c}_bq^{m-a-m_0}$, $N\geq1$. Furthermore, since $(q-b)q^{m-a-1}<(q-b+1)(q-1)q^{m-a-1}$, $N\leq q-b-1$. 

Assume that $|f_{\lambda_{b+N+1}}|\geq(N+1)q^{m-a-1}$. Then $$Nq^{m-a-1}+(q-b-N)(N+1)q^{m-a-1}\leq |f|= \widetilde{c}_bq^{m-a-m_0}$$ which gives $$Nq^{m_0-1}(q-N-b)\leq \widetilde{c}_b-q^{m_0-1}(q-b)<q^{m_0-1}.$$ This gives a contradiction since $1\leq N\leq q-b-1$. Furthermore the only possibility such that $|f_{\lambda_{b+N+1}}|=Nq^{m-a-1}$ is $N=1$ which is absurd since $f_{\lambda_{b+N+1}}$ is not a minimal weight codeword.

By Lemma \ref{2.15.1}, for all $b+1\leq i\leq N+b$, $g_{\lambda_{b+1}}=g_{\lambda_i}$. So, we can write for all $x=(x_1,\ldots,x_m)\in\F_q^m$ \begin{align*}f(x)&=\prod_{1\leq i\leq b}(x_1-\lambda_i)\left(g_{\lambda_{b+1}}(x_2,\ldots,x_m)+ \prod_{b+1\leq i\leq N+b}(x_1-\lambda_i)h(x)\right)\\&=\prod_{1\leq i\leq b}(x_1-\lambda_i)\left(\alpha f_{\lambda_{b+1}}(x_2,\ldots,x_m)+ \prod_{b+1\leq i\leq N+b}(x_1-\lambda_i)h(x)\right)\end{align*} where $h\in R_q(a(q-1)-N,m)$ and $\alpha\in\F_q^*$. 

Then, for all $(x_2,\ldots,x_m)\in\F_q^{m-1}$, $$f_{\lambda_{b+N+1}}(x_2,\ldots,x_m)=\beta f_{\lambda_{b+1}}(x_2,\ldots,x_m)+\gamma h_{\lambda_{b+N+1} }(x_2,\ldots,x_m).$$ We get a contradiction by Lemma \ref{2.14}.

Now, assume $n=1$, $n=2$, $n=b-2$ or $n=b-1$. 

Since $n\geq 1$, $f_{\lambda_1}=0$. So, $1-x_2^{q-1}$ divides $f_{\lambda_1}$. Then, there exists $k\in\{1,\ldots,q\}$ such that for all $i\leq k$, $1-x_2^{q-1}$ divides $f_{\lambda_i}$ and $(1-x_2^{q-1})$ does not divide $f_{\lambda_{k+1}}$. Since $S$ is not included in an hyperplane, $k\leq q-1$. For $i\geq n+1$, if $|f_{\lambda_i}|>(q-b+n)q^{m-a-2}$ then $$|f_{\lambda_i}|\geq w_2=\left\{\begin{array}{ll}q^{m-a-1}&\textrm{if $n=b-1$}\\q-b+n+1&\textrm{if $b\geq4$, $m=3$ and $n\neq b-1$}\\(q-b+n+1)(q-1)q^{m-a-3}&\textrm{otherwise}\end{array}\right..$$ Denote by $N:=\#\{i\geq n+1 :|f_{\lambda_i}|=(q-b+n)q^{m-a-2}\}$. Since $(q-n)w_2>\widetilde{c}_bq^{m-a-m_0}$, $N\geq1$. Since for $n=1$ or $n=b-1$, $(q-n)(q-b+n)q^{m-a-2}=(q-1)(q-b+1)q^{m-a-2}<|f|$, in these cases $N\leq q-n-1$. For $n=2$ or $n=b-2$, since  $(q-n)(q-b+n)q^{m-a-2}=(q-2)(q-b+2)q^{m-a-2}$, $N=q-n$.

Then, $|f_{\lambda_{n+1}}|=(q-b+n)q^{m-a-2}$ and $f_{\lambda_{n+1}}$ is a minimal weight codeword of $R_q(a(q-1)+b-n,m-1)$ so, by applying an affine transformation, we can assume $1-x_2^{q-1}$ divides $f_{\lambda_{n+1}}$. Thus, $k\geq n+1\geq 2$.

If $1\leq k\leq n+N-1$, then $|f_{\lambda_{k+1}}|=(q-b+n)q^{m-a-2}<(q-b+k)q^{m-a-2}$. If $n+N\leq k\leq q-1$ (since for $n=2$ or $n=b-2$, $n+N=q$, this case is possible only for $n=1$ or $n=b-1$), assume $|f_{\lambda_{k+1}}|\geq (q-b+k)q^{m-a-2}$. Then, $$|f|\geq N(q-b+n)q^{m-a-2}+(k-n-N)w_2+(q-k)(q-b+k)q^{m-a-2}. $$ Since $|f|=\widetilde{c}_bq^{m-a-m_0}$, $c_b<(q-b+1)-1$, $1\leq N\leq q-n-1$ and $n+N\leq k\leq q-1$, we get a contradiction.

Since for all $n\leq i\leq k$, $1-x_2^{q-1}$ divides $f_{\lambda_i}$, it divides $g_{\lambda_i}$ too. Then we can write for all $x=(x_1,x_2,\ldots,x_m)\in\F_q^{m}$ \begin{align*}f(x)&=\prod_{1\leq i\leq n}(x_1-\lambda_i)\left(\prod_{n+1\leq i\leq k}(x_1-\lambda_i)h(x_1,x_2,x_3,\ldots,x_m)\right.\\&\hspace{2cm}+(1-x_2^{q-1})l(x_1,x_3,\ldots,x_m)\Bigg)\end{align*} with $\deg(h)\leq a(q-1)+b-k$ and $l\in R_q((a-1)(q-1)+b-n,m-1)$. Then for all $(x_2,\ldots,x_m)\in\F_q^{m-1}$, $$f_{\lambda_{k+1}}(x_2, \ldots,x_m)=\alpha h_{\lambda_{k+1}}(x_2,\ldots,x_m)+\beta(1-x_2^{q-1})l_{\lambda_{k+1}}(x_3,\ldots,x_m).$$ 
 Thus, we get a contradiction by Lemma \ref{3.5} since $k\geq2$ and $|f_{\lambda_{k+1}}|<(q-b+k)q^{m-a-2}$.\end{preuve}

\begin{theoreme}\label{m3}For $q\geq 7$, $m\geq2$, $0\leq a\leq m-2$, $4\leq b\leq q-2$, up to affine transformation, if $c_b<(q-b+1)q-1$, the third weight codeword of $R_q(a(q-1)+b,m)$ are of the form : $$\forall x=(x_1,\ldots,x_m)\in\F_q^m,\quad f(x)=\prod_{i=1}^a(1-x_1^{q-1})g(x_{a+1},x_{a+2})$$ where $g\in R_q(b,2)$ is such that $|g|=c_b$. 
\end{theoreme}

\begin{theoreme}\label{m33}For $q\geq 7$, $m\geq3$, $0\leq a\leq m-3$, up to affine transformation, the third weight codeword of $R_q(a(q-1)+3,m)$ are of the form : $$\forall x=(x_1,\ldots,x_m)\in\F_q^m,\quad f(x)=\prod_{i=1}^a(1-x_1^{q-1})x_{a+1}x_{a+2}x_{a+3}$$. 
\end{theoreme}

\begin{preuve}Let $f\in R_q(a(q-1)+b,m)$ such that $|f|=\widetilde{c}_bq^{m-a-m_0}$. We denote by $S$ the support of $f$.
If $a\geq1$ then by Lemma \ref{suphyp} or Lemma \ref{suphyp2}, $S$ is included in an hyperplane. By applying an affine transformation, we can assume $S$ is included in the hyperplane of equation $x_1=0$. Then by Lemma \ref{DGMW1}, for all $x=(x_1,\ldots,x_m)\in\F_q^m$, we can write $$f(x)=(1-x_1^{q-1})g_1(x_2,\ldots,x_m)$$ with $g_1\in R_q((a-1)(q-1)+b,m-1)$. Since $|g_1|=|f|=\widetilde{c}_bq^{m-a-m_0}$, we can iterate this process. So, for all $x=(x_1,\ldots,x_m)\in\F_q^m$, $$f(x)=\prod_{i=1}^a(1-x_i^{q-1})g_a(x_{a+1},\ldots,x_m)$$ with $g_a\in R_q(b,m-a)$ and $|g_a|=|f|=\widetilde{c}_bq^{m-a-m_0}$.

Then $g_a$ fulfils the conditions of Proposition \ref{case1} or Proposition \ref{case2} and we get the result. \end{preuve}

\section{Conclusions}

In this paper, we describe the third weight of generalized Reed-Muller codes for small $b\geq2$. More precisely, for $4\leq b< \frac{q+3}{2}$, $c_b\leq(q-b+2)(q-2)<(q-b+1)q-1$. Then from the results of Section \ref{cb}, Theorem \ref{w3}, Theorem \ref{w33}, Theorem \ref{m3} and Theorem \ref{m33}, we know the third weight and the third weight codewords of generalized Reed-Muller codes for $3\leq b<\frac{q+3}{4}$. If $\frac{q+3}{4}\leq b<\frac{q+3}{2}$, we know the third weight and the third weight codewords of generalized Reed-Muller codes up to the third weight and the third weight codewords of $R_q(b,2)$. If $b=2$ or $b=\frac{q+3}{2}$, $c_b\leq (q-b+2)(q-2)=(q-b+1)-1$, so for $b=2$, we know the third weight and for $b=\frac{q+3}{2}$, we know the third weight up to the third weight of $R_q(b,2)$ but we do not know the form of the third weight codewords. Finally, if $b\geq \frac{q}{2}+2$, $c_b\leq (q-b+1)q$ we are not in the conditions of application of any of the previous theorems. This upper bound on $c_b$ is the minimum weight of $R_q(b-1,2)$. Either we can find $f$ a polynomial (of degree $b$) such that $|f|\leq(q-b+1)q-1$ and we will have a better upper bound on the third weight of $R_q(a(q-1)+b,m)$ for $0\leq a\leq m-2$ and $b\geq \frac{q}{2}+2$ or $c_b=(q-b+1)q$. In this last case several questions arise : is the third weight of $R_q(b,2)$ reached only by minimal weight codeword of $R_q(b-1,2)$? Is the third weight of $R_q(a(q-1)+b,m)$ also the minimal weight of $R_q(a(q-1)+b-1,m)$?

\appendix
\section{Appendix : Blocks of hyperplane arrangements}

\subsection{Basic facts}
Let $1\leq k\leq m$ and $1\leq d_i\leq q-1$. We say that $\mathcal{B}$ is an hyperplane arrangement of type $(k,d_1,\ldots,d_k)$ if $f_1,\ldots,f_k$ are k independent linear forms over $\F_q^m$ and $\mathcal{B}$ is composed of $k$ blocks of $d_i$ parallel hyperplanes of equation $f_i(x)=u_i,j$ where $1\leq i\leq k$, $1\leq j\leq d_i$, $u_{i,j}\in\F_q$ and if $k\neq j$, $u_{i,j}\neq u_{i,k}$.

We denote by $\mathcal{L}_d$ the set of hyperplane arrangements of type $(k,d_1,\ldots,d_k)$, $1\leq k\leq m$ and $1\leq d_i\leq q-1$ such that $\displaystyle\sum_{i=1}^kd_i\leq d$. 

For $\mathcal{A}\in\mathcal{L}_d$, we denote by $N(\mathcal{A})=\#\displaystyle\bigcup_{H\in\mathcal{A}}H$.

We can find the following theorem in \cite{MR2592428}

\begin{theoreme}Let $\mathcal{A}\in\mathcal{L}_d$ then, $N(\mathcal{A})=q^m-q^{m-k}\displaystyle\prod_{i=1}^k(q-d_i)$.\end{theoreme}

\subsection{Maximal and second weight configurations}

We write $d=t(q-1)+s$ where $0\leq t\leq m-1$ and $0\leq s\leq q-2$. We say that $\mathcal{A}\in\mathcal{L}_d$ is a maximal configuration if $N(\mathcal{A})=(q-s)q^{m-t-1}$.

It is proved in \cite{delsarte_poids_min} (see also \cite{Leducq2012581}) that a maximal configuration is :

$T_{max}$ : $t$ blocks of $q-1$ hyperplanes and 1 block of $s$ hyperplanes.

A second weight configuration is $\mathcal{A}\in\mathcal{L}_d$ such that :

\begin{enumerate}
\item If $q\geq4$
\begin{itemize}
\item For $0\leq t\leq m-2$, $2\leq s\leq q-2$, $N(\mathcal{A})=q^m-(q-s+1)(q-1)q^{m-t-2}$.
\item For $0\leq t\leq m-1$, $s=1$, $N(\mathcal{A})=q^m-q^{m-t}$
\item For $1\leq t \leq m-1$, $s=0$, $N(\mathcal{A})=q^m-2(q-1)q^{m-t-1}$ 
\end{itemize}
\item If $q=3$
\begin{itemize}
\item For $1\leq t\leq m-1$, $s=0$, $N(\mathcal{A})=3^m-4.3^{m-t-1}$
\item For $1\leq t\leq m-2$, $s=1$, $N(\mathcal{A})=3^m-8. 3^{m-t-2}$.
\item For $t=m-1$, $s=1$, $N(\mathcal{A})=3^m-3$.
\end{itemize}
\end{enumerate}

It is proved in \cite{raey} and \cite{MR2592428} that a second weight configuration is among the following ones :

\begin{enumerate}\item $T_1$ : $t-1$ blocks with $q-1$ hyperplanes, 1 block with $q-2$ hyperplanes, 1 block with $s+1$ hyperplanes (for $1\leq t\leq m-1$, $0\leq s\leq q-3$)
\item $T_2$ : $t-1$ blocks with $q-1$ hyperplanes, 1 block with $q-2$ hyperplanes, 1 block with $s$ hyperplanes, 1 block with 1 hyperplane (for $1\leq t\leq m-2$, $1\leq s\leq q-2$) 
\item $T_3$ : $t$ blocks with $q-1$ hyperplanes, $1$ block with $s-1$ hyperplanes, 1 block with 1 hyperplane (for $0\leq t\leq m-2$, $2\leq s\leq q-2$) 
\item $T_4$ : $t$ blocks with $q-1$ hyperplanes (for $0\leq t\leq m-1$).
\end{enumerate}

More precisely,

\begin{enumerate}
\item if $q\geq4$,
\begin{itemize}
\item for $0\leq t\leq m-2$, $2\leq s\leq q-2$, a second weight configuration is $T_3$,
\item for $0\leq t\leq m-1$, $s=1$, a second weight configuration is $T_4$,
\item for $1\leq t \leq m-1$, $s=0$, a second weight configuration is $T_1$,
\end{itemize}
\item if $q=3$,
\begin{itemize}
\item for $1\leq t\leq m-1$, $s=0$, a second weight configuration is $T_1$,
\item for $1\leq t\leq m-2$, $s=1$, a second weight configuration is $T_2$.
\item for $t=m-1$, $s=1$, a second weight configuration is $T_4$.
\end{itemize}
\end{enumerate}

\subsection{Modification of the second weight configurations}

\subsubsection{Modification of $T_1$}

Since $T_1$ is the second weight configuration in the case $s=0$, we consider here only this case. We consider then 5 modifications on this configuration :

\begin{enumerate}
\item $T_1^{(a)}$ : $t-1$ blocks with $q-1$ hyperplanes, 1 block with $q-3$ hyperplanes, 1 block with 2 hyperplanes (for $1\leq t\leq m-1$). 

We obtain this configuration by replacing one of the hyperplanes of the block with $q-2$ hyperplanes in $T_1$ by one hyperplane of the block with 1 hyperplane. 
$$N(T_1^{(a)})=q^m-3(q-2)q^{m-t-1}.$$
\item $T_1^{(b)}$ : $t-1$ blocks with $q-1$ hyperplanes, 1 block with $q-3$ hyperplanes, 2 blocks with one hyperplanes (for $1\leq t\leq m-2$).

We obtain this configuration by replacing one of the hyperplanes of the block with $q-2$ hyperplanes in $T_1$ by a new block with one hyperplane.
$$N(T_1^{(b)})=q^m-3(q-1)^2q^{m-t-2}.$$
\item $T_1^{(c)}$ : $t-2$ blocks with $q-1$ hyperplanes, 2 blocks with $q-2$ hyperplanes, 1 block with 2 hyperplanes (for $2\leq t\leq m-1$).

We obtain this configuration by replacing one of the hyperplanes of a complete block in $T_1$ by one hyperplane of the block with 1 hyperplane.
$$N(T_1^{(c)})=q^m-4(q-2)q^{m-t-1}.$$
\item $T_1^{(d)}$ : $t-2$ blocks with $q-1$ hyperplanes, 2 blocks with $q-2$ hyperplanes, 2 blocks with 1 hyperplane (for $2\leq t\leq m-2$).

We obtain this configuration by replacing one of the hyperplanes of a complete block in $T_1$ by a new block with one hyperplane.
$$N(T_1^{(d)})=q^m-4(q-1)^2q^{m-t-2}.$$
\item $T_1^{(e)}$ : $t-1$ blocks with $q-1$ hyperplanes, 1 block with $q-2$ hyperplanes (for $1\leq t\leq m-1$).

We obtain this configuration by removing the bloc with 1 hyperplane in $T_1$.
$$N(T_1^{(e)})=q^m-2q^{m-t}.$$
\end{enumerate}

In this case, we denote by $N'_3=\max(N(T_1^{(a)}),N(T_1^{(b)}),N(T_1^{(c)}),N(T_1^{(d)}),N(T_1^{(e)}))$. If a configuration in is not defined, is a maximal configuration or a second weight configuration, we do not consider it in the max.

\begin{remarque}\begin{itemize}\item If $q=3$, $T_1^{(a)}$ is a maximal configuration, $T_1^{(b)}$ and $T_1^{(c)}$ are $T_1$ configuration. So, the only configuration above that we have to consider are $T_1^{(d)}$ and $T_1^{(e)}$.
\item If $q=4$, $T_1^{(a)}=T_1$ and we do not consider this configuration in this case.
\end{itemize}\end{remarque}

\begin{proposition}\label{1} $$N'_3=\left\{\begin{array}{ll}N(T_1^{(e)})&\textrm{$\begin{array}{l}\textrm{if $q\geq7$}\\\textrm{or $q=4$ and $t=m-1$}\\\textrm{or $q=3$ and $t=1$ or $t=m-1$} \end{array}$}\\N(T_1^{(b)})&\textrm{if $q=4$ and $t\leq m-2$}\\N(T_1^{(a)})&\textrm{if $q=5$}\\N(T_1^{(d)})&\textrm{if $q=3$ and $m-2\geq t\geq2$}\end{array}\right..$$\end{proposition}

\begin{remarque}If $q=4$ and $t=m-1$, $N(T_1^{(e)})=N(T_1^{(c)})$.\end{remarque}

\begin{preuve}\begin{align*}N(T_1^{(a)})-N(T_1^{(e)})&=q^m-3(q-2)q^{m-t-1}-q^m+2q^{m-t}
\\&=q^{m-t-1}(-q+6)\end{align*}
If $q=3$ or $q=4$, we do not consider $N(T_1^{(a)})$ to determine $N'_3$. Otherwise, if $q=5$ then, $N(T_1^{(a)})>N(T_1^{(e)})$ and if $q\geq 7$ then, $N(T_1^{(e)})>N(T_1^{(a)})$.

\begin{align*}N(T_1^{(b)})-N(T_1^{(e)})&=q^m-3(q-1)^2q^{m-t-2}-q^m+2q^{m-t}\\&=q^{m-t-2}(-3(q-1)^2+2q^2)\\&=q^{m-t-2}(-q^2+6q-3)\end{align*}
If $q=3$ we do not consider $N(T_1^{(b)})$ to determine $N'_3$. Otherwise, if $q\leq5$ then, $N(T_1^{(b)})>N(T_1^{(e)})$ and if $q\geq 7$ then, $N(T_1^{(e)})>N(T_1^{(b)})$.

\begin{align*}N(T_1^{(c)})-N(T_1^{(e)})&=q^m-4(q-2)q^{m-t-1}-q^m+2q^{m-t}\\&=q^{m-t-1}(-2q+8)\end{align*}
If $q=3$ we do not consider $N(T_1^{(c)})$ to determine $N'_3$. Otherwise, if $q=4$ then, $N(T_1^{(c)})=N(T_1^{(e)})$ and if $q\geq 5$ then, $N(T_1^{(e)})>N(T_1^{(c)})$.

\begin{align*}N(T_1^{(d)})-N(T_1^{(e)})&=q^m-4(q-1)^2q^{m-t-2}-q^m+2q^{m-t}\\&=q^{m-t-2}(2q^2-4(q-1)^2)\\&=q^{m-t-2}(-2q^2+8q-4)\end{align*}
If $q=3$ and $t=1$ or $t=m-1$ the only configuration we consider to determine $N'_3$ is $T_1^{(e)}$. If $q=3$ and $m-2\geq t\geq 2$, $N'_3= N(T_1^{(d)})$. If $q\geq 4$ then, $N(T_1^{(e)})>N(T_1^{(d)})$.

Now to complete the proof we only have to compare $N(T_1^{(a)})$ and $N(T_1^{(b)})$ in the case where $q=5$.
\begin{align*}N(T_1^{(a)})-N(T_1^{(b)})&=5^m-9.5^{m-t-1}-5^m+48.5^{m-t-2}\\&=5^{m-t-2}(48-9.5)\\&=3.5^{m-t-2}>0\end{align*}
\end{preuve}

\subsubsection{Modification of $T_2$}

Since $T_2$ is the second weight configuration in the case where $q=3$, $1\leq t\leq m-2$ and $s=1$, we consider here only this case. We consider then 1 modification on this configuration :
\begin{enumerate}
\item $T_2^{(a)}$ : $t-2$ blocks with 2 hyperplanes and 5 blocks with 1 hyperplane (for $2\leq t\leq m-3$).

We obtain this configuration by replacing one of the hyperplane of a complete block of $T_2$ by a new block with one hyperplane.
$$N(T_2^{(a)})=3^m-32. 3^{m-t-3}.$$

\end{enumerate}

In this case, we denote by $N'_3=\max(N(T_4), N(T_2^{(a)}))$. If a configuration in max is not defined, is a maximal configuration or a second weight configuration, we do not consider it in the max.

\begin{proposition}$N'_3=N(T_4)$.\end{proposition}

\begin{preuve}\begin{align*}N(T_2^{(a)})-N(T_4)&=3^m-32. 3^{m-t-3}-3^m+3^{m-t}\\&=3^{m-t-3}(-32+27)<0\end{align*}

So, $N'_3=N(T_4)$.\end{preuve}

\subsubsection{Modification of $T_3$}

Since $T_3$ is the second weight configuration in the case where $q\geq 4$, $0\leq t\leq m-2$ and $2\leq s\leq q-2$, we only consider this case. We consider then 5 modifications of this configuration :
\begin{enumerate}
\item $T_3^{(a)}$ : $t$ blocks with $q-1$ hyperplanes, 2 blocks with 1 hyperplane and 1 block with $s-2$ hyperplanes (for $0\leq t\leq m-3$ and $3\leq s\leq q-2$).

we obtain this configuration by replacing 1 hyperplane from the block with $s-1$ hyperplanes in $T_3$ by a new block with one hyperplane.
$$N(T_3^{(a)})=q^m-q^{m-t-3}(q-1)^2(q-s+2).$$
\item $T_3^{(b)}$ : $t-1$ blocks with $q-1$ hyperplanes, 1 block with $q-2$ hyperplanes, 1 block with $s-1$ hyperplanes and 1 block with 2 hyperplanes (for $1\leq t\leq m-2$ and $2\leq s\leq q-2$). 

We obtain this configuration by replacing one hyperplane from a complete block of $T_3$ by an hyperplane from the block with 1 hyperplane in $T_3$.
$$N(T_3^{(b)})=q^m-q^{m-t-2}2(q-2)(q-s+1).$$
\item $T_3^{(c)}$ : $t-1$ blocks with $q-1$ hyperplanes, 1 block with $q-2$ hyperplanes, 2 blocks with 1 hyperplane and 1 block with $s-1$ hyperplanes (for $1\leq t\leq m-3$ and $2\leq s\leq q-2$). 

We obtain this configuration by replacing 1 hyperplane from a complete block of $T_3$ by an hyperplane from a new block.
$$N(T_3^{(c)})=q^m-q^{m-t-3}2(q-1)^2(q-s+1).$$
\item $T_3^{(d)}$ : $t$ blocks with $q-1$ hyperplanes, 1 block with $s-2$ hyperplanes, 1 block with 2 hyperplanes (for $0\leq t\leq m-2$ and $4\leq s\leq q-2$).

We obtain this configuration by replacing 1 hyperplane from the block with $s-1$ hyperplanes in $T_3$ by 1 hyperplane of the block with 1 hyperplane in $T_3$.
$$N(T_3^{(d)})=q^m-q^{m-t-2}(q-2)(q-s+2).$$
\item $T_3^{(e)}$ : $t$ blocks with $(q-1)$ hyperplanes and 1 block with $s-1$ hyperplanes (for $0\leq t\leq m-2$ and $2\leq s\leq q-2$).

We obtain this configuration by removing the block with 1 hyperplane in $T_3$.
$$N(T_3^{(e)})=q^m-q^{m-t-1}(q-s+1).$$
\end{enumerate}

In this case, we denote by $$N'_3=\max(N(T_1), N(T_2), N(T_3^{(a)}),N(T_3^{(b)}),N(T_3^{(c)}),N(T_3^{(d)}),N(T_3^{(e)})).$$ If a configuration in max is not defined, is a maximal or a second weight configuration, we do not consider it in the max.

\begin{proposition}$$N'_3=\left\{\begin{array}{ll}N(T_3^{(d)})& \textrm{if $q\geq7$, $0\leq t\leq m-2$ and $4\leq s\leq\lfloor\frac{q}{2}+2\rfloor$}\\&\\N(T_3^{(e)})&\textrm{$\begin{array}{l}\textrm{if $q\geq8$, $0\leq t\leq m-2$ and $\lceil\frac{q}{2}+2\rceil\leq s\leq q-2$}\\ \textrm{or $q\geq4$, $ 0\leq t\leq m-2$ and $s=2$}\\\textrm{or $q\geq5$, $t=m-2$ and $s=3$}\end{array}$}
\\&\\N(T_3^{(a)})&\textrm{if $q\geq5$, $0\leq t\leq m-3$, $s=3$} \end{array}\right..$$\end{proposition}

\begin{remarque}\begin{itemize}\item If $q$ is even, $s=\frac{q}{2}+2$, $N(T_3^{(d)})=N(T_3^{(e)})$. 
\item If $q=5$, $t=m-2$ and $s=2$, $N(T_1)=N(T_3^{(e)})$.\end{itemize}\end{remarque}
\begin{preuve}\begin{align*}N(T_1)-N(T_3^{(e)})&=q^m-2q^{m-t-1}(q-s-1)-q^m+q^{m-t-1}(q-s+1)
\\&=q^{m-t-2}(-q+s+3).\end{align*}

So, if $s\leq q-4$, $N(T_1)<N(T_3^{(e)})$, if $s=q-3$, $N(T_1)=N(T_3^{(e)})$,  if $s=q-2$, $T_1$ is a minimal weight configuration.

\begin{align*}N(T_2)-N(T_3^{(e)})&=q^m-2q^{m-t-2}(q-1)(q-s)-q^m+q^{m-t-1}(q-s+1)
\\&=q^{m-t-2}(-2(q-1)(q-s)+q(q-s+1))\\&=q^{m-t-2}(-q(q-3-s)-2s).\end{align*}
So, if $s\leq q-3$, $N(T_2)<N(T_3^{(e)})$ and if $s=q-2$, $N(T_2)\leq N(T_3^{(e)})$.

\begin{align*}N(T_3^{(a)})-N(T_3^{(e)})&=q^m-q^{m-t-3}(q-1)^2(q-s+2)-q^m+q^{m-t-1}(q-s+1)
\\&=q^{m-t-3}(-(q-1)^2(q-s+2)+q^2(q-s+1))\\&=q^{m-t-3}(q^2-2sq+3q+s-2)\end{align*}
So, if $s>\frac{q^2+3q-2}{2q-1}$, $N(T_3^{(a)})<N(T_3^{(e)})$ and if $s<\frac{q^2+3q-2}{2q-1}$, $N(T_3^{(a)})>N(T_3^{(e)})$.

\begin{align*}N(T_3^{(b)})-N(T_3^{(e)})&=q^m-q^{m-t-2}2(q-2)(q-s+1)-q^m+q^{m-t-1}(q-s+1)
\\&=q^{m-t-2}(q-s+1)(-q+4)\end{align*}
So, if $q\geq 5$, $N(T_3^{(b)})<N(T_3^{(e)})$ and if $q=4$, $N(T_3^{(b)})=N(T_3^{(e)})$.

\begin{align*}N(T_3^{(c)})-N(T_3^{(e)})&=q^m-2(q-1)^2(q-s+1)q^{m-t-3}-q^m+q^{m-t-1}(q-s+1)
\\&=q^{m-t-3}(q-s+1)(-2(q-1)^2+q^2)\\&=q^{m-t-3}(q-s+1)(-q(q-4)-2)<0\end{align*}

So, $N(T_3^{(c)})<N(T_3^{(e)})$.

\begin{align*}N(T_3^{(e)})-N(T_3^{(d)})&=q^m-q^{m-t-1}(q-s+1)-q^m+q^{m-t-2}(q-2)(q-s+2)
\\&=q^{m-t-2}((q-2)(q-s+2)-q(q-s+1))\\&=q^{m-t-2}(2s-4-q)\end{align*}
So, if $s<\frac{q}{2}+2$, then $N(T_3^{(e)})<N(T_3^{(d)})$, if $s>\frac{q}{2}+2$, $N(T_3^{(e)})>N(T_3^{(d)})$ and if $s=\frac{q}{2}+2$, $N(T_3^{(e)})=N(T_3^{(d)})$.\\

Now, $\frac{q^2+3q-2}{2q-1}=\frac{q}{2}+\frac{7}{4}-\frac{1}{4}.\frac{1}{2q-1}$ and $\frac{3}{2}<\frac{7}{4}-\frac{1}{4}.\frac{1}{2q-1}<2$. Then, since $s$ is an integer, $s<\frac{q}{2}+2\Leftrightarrow s<\frac{q^2+3q-2}{2q-1}$. So, if $s>\frac{q}{2}+2$ then $s>\frac{q^2+3q-2}{2q-1}$ and $N(T_3^{(e)})>N(T_3^{(a)})$. If $s<\frac{q}{2}+2$ we need to compare $N(T_3^{(a)})$ and $N(T_3^{(d)})$.

\begin{align*}N(T_3^{(a)})-N(T_3^{(d)})&=q^m-q^{m-t-3}(q-1)^2(q-s+2)-q^m+q^{m-t-2}(q-2)(q-s+2)
\\&=-q^{m-t-3}(q-s+2)\end{align*}
So, $N(T_3^{(a)})<N(T_3^{(d)})$.
\end{preuve}

\subsubsection{Modification of $T_4$}

Since $T_4$ is the second weight configuration in the case where $q\geq 4$, $0\leq t\leq m-1$ and $s=1$ or where $q=3$, $t=m-1$ and $s=1$, we consider here only these cases. We consider 1 modification for this configuration : 

\begin{enumerate}
\item $T_4^{(a)}$ : $t-1$ blocks with $q-1$ hyperplanes, 1 block with $q-2$ hyperplanes, 1 block with 1 hyperplane (for $1\leq t\leq m-1$).

We obtain this configuration by replacing 1 hyperplane from a complete block by an hyperplane from a new block.
$$N(T_4^{(a)})=q^m-2(q-1)q^{m-t-1}.$$
\end{enumerate}

In this case, we denote by $N'_3=\max(N(T_1), N(T_2), N(T_4^{(a)}))$.

If a configuration in $\max$ is not defined or is a maximal configuration or a second weight configuration, we do not consider it in the $\max$.

\begin{proposition}\label{4}$$N'_3=\left\{\begin{array}{ll}N(T_1)&\textrm{if $q\geq 5$}\\N(T_2)&\textrm{if $q=4$ and $1\leq t\leq m-2$}\\N(T_4^{(a)})&\textrm{if $q=3$ or $q=4$ and $t=m-1$}\end{array}\right..$$\end{proposition}

\begin{preuve}\begin{align*}N(T_1)-N(T_4^{(a)})&=q^m-2(q-2)q^{m-t-1}-q^m+2(q-1)q^{m-t-1}\\&=2q^{m-t-1}>0\end{align*}

So, $N(T_1)>N(T_4^{(a)})$.

\begin{align*}N(T_2)-N(T_4^{(a)})&=q^m-2(q-1)^2q^{m-t-2}-q^m+2(q-1)q^{m-t-1}\\&=2(q-1)q^{m-t-2}>0\end{align*}

So, $N(T_2)>N(T_4^{(a)})$.

\begin{align*}N(T_2)-N(T_1)&=q^m-2(q-1)^2q^{m-t-2}-q^m+2(q-2)q^{m-t-1}\\&=-2q^{m-t-2}<0\end{align*}

So, $N(T_1)>N(T_2)$.
\end{preuve}

\subsection{Third weight for hyperplane arrangements}

\begin{proposition}\label{5}For $q\geq 3$, $1\leq t\leq m-1$ and $s=0$, if $\mathcal{B}\in\mathcal{L}_d\setminus\{T_1^{(a)},T_1^{(b)},T_1^{(c)},T_1^{(d)},T_1^{(e)}\}$ and $\mathcal{B}$ is not a maximal or second weight configuration then, $$N(\mathcal{B})<N'_3.$$  \end{proposition}

\begin{preuve}Let $k$ and $d_1,\ldots,d_k$ defining an hyperplane arrangement $\mathcal{B}\in\mathcal{L}_d\setminus\{T_{max},T_1,T_1^{(a)},T_1^{(b)},T_1^{(c)},T_1^{(d)},T_1^{(e)}\}$. Then $$N(\mathcal{B})=q^m-q^{m-k}=\prod_{i=1}^k(q-d_i).$$
Let $d'=\displaystyle\sum_{i=1}^kd_i=t'(q-1)+s'$, $0\leq s'\leq q-2$.
\begin{itemize}\item Assume there exists two distinct $1\leq i_1\leq k$ and $1\leq i_2\leq k$ such that $$1\leq d_{i_1}\leq d_{i_2}\leq q-2.$$
Then let replace an hyperplane of the block $i_1$ in $\mathcal{B}$ by a new hyperplane of the block $i_2$ in $\mathcal{B}$. We denote by $\mathcal{B}'$ this new hyperplane arrangement. Then,
\begin{align*}N(\mathcal{B}')-N(\mathcal{B})&=q^m-q^{m-k}\prod_{i\neq i_1,i_2}(q-d_i)(q-d_{i_1}+1)(q-d_{i_2}-1)\\&\hspace{2cm}-q^m+q^{m-k}\prod_{i=1}^k(q-d_i)
\\&=q^{m-k}\prod_{i\neq i_1,i_2}(q-d_i)(d_{i_2}-d_{i_1}+1)>0.\end{align*}
Since $\mathcal{B}$ is not a $T_1$ configuration, $\mathcal{B'}$ is not a maximal configuration. Since $\mathcal{B}$ is not a configuration of type $T_1^{(a)}$, $T_1^{(b)}$, $T_1^{(c)}$ or $T_1^{(d)}$, $\mathcal{B}'$ is not a $T_1$ configuration. So, in this case, $N(\mathcal{B})<N_3'$.
\item If there exists $1\leq i_1\leq k$ such that all $d_i$ but $d_{i_1}$ are 0 or $q-1$ then, $\mathcal{B}$ is composed of $t'$ blocks with $q-1$ hyperplanes and 1 block with $s'$ hyperplanes. Since $d'\leq d$, necessarily $t'<t\leq m-1$. So we can add 1 hyperplane in a new direction. We denote by $\mathcal{B}'$ this new configuration. Then, 
\begin{align*}N(\mathcal{B}')-N(\mathcal{B})&=q^m-q^{m-k-1}\prod_{i=1}^k(q-d_i)(q-1)-q^m+q^{m-k}\prod_{i=1}^k(q-d_i)
\\&=q^{m-k-1}\prod_{i=1}^k(q-d_i)>0.\end{align*}
By definition of $\mathcal{B}'$, $\mathcal{B'}$ is not a maximal configuration. Since $\mathcal{B}$ is not a $T_1^{(e)}$ configuration, $\mathcal{B}'$ is not a $T_1$ configuration. So, in this case, $N(\mathcal{B})<N_3'$.
\item Assume for all $1\leq i\leq k$, $d_i$ is either 0 or $q-1$. Then, $\mathcal{B}$ is composed of $t'$ blocks of $q-1$ hyperplanes and $s'=0$. Since $t'\leq t\leq m-1$, we can add a new hyperplane in a new direction. We denote by $\mathcal{B}'$ this new configuration and we have $N(\mathcal{B}')>N(\mathcal{B})$. By definition, $\mathcal{B'}$ is neither a maximal configuration nor a $T_1$ configuration. So, in this case, $N(\mathcal{B})<N_3'$.
\end{itemize}\end{preuve}

\begin{proposition}For $q= 3$, $1\leq t\leq m-2$ and $s=1$, if $\mathcal{B}\in\mathcal{L}_d\setminus\{T_4,T_2^{(a)}\}$ and $\mathcal{B}$ is not a maximal or second weight configuration then, $$N(\mathcal{B})<N'_3.$$  \end{proposition}

\begin{preuve}Let $k$ and $d_1,\ldots,d_k$ defining an hyperplane arrangement $\mathcal{B}\in\mathcal{L}_d\setminus\{T_{max},T_2,T_4,T_2^{(a)}\}$ and $d'=\displaystyle\sum_{i=1}^kd_i=t'(q-1)+s'$, $0\leq s'\leq q-2$.
\begin{itemize}\item Assume there exists two distinct $1\leq i_1\leq k$ and $1\leq i_2\leq k$ such that $$1\leq d_{i_1}\leq d_{i_2}\leq q-2.$$
Then let replace an hyperplane of the block $i_1$ in $\mathcal{B}$ by a new hyperplane of the block $i_2$ in $\mathcal{B}$. We denote by $\mathcal{B}'$ this new hyperplane arrangement. Then, $N(\mathcal{B}')>N(\mathcal{B})$.
Since $\mathcal{B}$ is not a $T_2$ configuration, $\mathcal{B'}$ is not a maximal configuration. Since $\mathcal{B}$ is not a configuration of type $T_2^{(a)}$, $\mathcal{B}'$ is not a $T_2$ configuration. So, in this case, $N(\mathcal{B})<N_3'$.
\item Assume there exists $1\leq i_1\leq k$ such that all $d_i$ but $d_{i_1}$ are 0 or $q-1$. Then, since $t'\leq t\leq m-2$, we can add 1 hyperplane in a new direction. We denote by $\mathcal{B}'$ this new configuration and we have 
$N(\mathcal{B}')>N(\mathcal{B})$. By definition, $\mathcal{B}'$ is neither a maximal configuration nor a $T_2$ configuration. So, in this case, $N(\mathcal{B})<N_3'$.
\item Assume for all $1\leq i\leq k$, $d_i$ is either 0 or $q-1$. Then, we can add a new hyperplane in a new direction. We denote by $\mathcal{B}'$ this new configuration and  we have $N(\mathcal{B}')>N(\mathcal{B})$. Since $\mathcal{B}$ is not a $T_4$ configuration, $\mathcal{B}'$ is not a maximal configuration. By definition, $\mathcal{B}'$ is not a $T_2$ configuration. So, in this case, $N(\mathcal{B})<N_3'$.
\end{itemize}\end{preuve}

\begin{proposition}For $q\geq 4$, $0\leq t\leq m-2$ and $2\leq s\leq q-2$, if $\mathcal{B}\in\mathcal{L}_d\setminus\{T_1,T_2,T_3^{(a)},T_3^{(b)},T_3^{(c)},T_3^{(d)},T_3^{(e)}\}$ and $\mathcal{B}$ is not a maximal or second weight configuration then, $$N(\mathcal{B})<N'_3.$$  \end{proposition}

\begin{preuve}Let $k$ and $d_1,\ldots,d_k$ defining an hyperplane arrangement $\mathcal{B}\in\mathcal{L}_d\setminus\{T_{\max},T_3,T_1,T_2,T_3^{(a)},T_3^{(b)},T_3^{(c)},T_3^{(d)},T_3^{(e)}\}$ and $d'=\displaystyle\sum_{i=1}^kd_i=t'(q-1)+s'$, $0\leq s'\leq q-2$.
\begin{itemize}\item Assume there exists two distinct $1\leq i_1\leq k$ and $1\leq i_2\leq k$ such that $$1\leq d_{i_1}\leq d_{i_2}\leq q-2.$$
Then let replace an hyperplane of the block $i_1$ in $\mathcal{B}$ by a new hyperplane of the block $i_2$ in $\mathcal{B}$. We denote by $\mathcal{B}'$ this new hyperplane arrangement. Then, $N(\mathcal{B}')>N(\mathcal{B})$. Since $\mathcal{B}$ is not a configuration of type $T_1$, $T_2$ or $T_3$, $\mathcal{B'}$ is not a maximal configuration. Since $\mathcal{B}$ is not a configuration of type $T_3^{(a)}$, $T_3^{(b)}$, $T_3^{(c)}$ or $T_3^{(d)}$, $\mathcal{B}'$ is not a $T_3$ configuration. So, in this case, $N(\mathcal{B})<N_3'$.
\item Assume there exists $1\leq i_1\leq k$ such that all $d_i$ but $d_{i_1}$ are 0 or $q-1$. Then, since $t'\leq t\leq m-2$, we can add 1 hyperplane in a new direction. We denote by $\mathcal{B}'$ this new configuration and we have 
$N(\mathcal{B}')>N(\mathcal{B})$. By definition, $\mathcal{B}'$ is not a maximal configuration. Since $\mathcal{B}$ is not a $T_3^{(e)}$ configuration, $\mathcal{B}'$ is not a $T_3$ configuration. So, in this case, $N(\mathcal{B})<N_3'$.
\item Assume for all $1\leq i\leq k$, $d_i$ is either 0 or $q-1$. Then we can add a new hyperplane in a new direction. We denote by $\mathcal{B}'$ this new configuration and  we have $N(\mathcal{B}')>N(\mathcal{B})$. By definition, $\mathcal{B}'$ is neither a maximal configuration nor a $T_3$ configuration. So, in this case, $N(\mathcal{B})<N_3'$.
\end{itemize}\end{preuve}

\begin{proposition}\label{6}For $q\geq 4$, $0\leq t\leq m-1$ and $s=1$ or $q=3$, $t=m-1$ and $s=1$, if $\mathcal{B}\in\mathcal{L}_d\setminus\{T_1,T_2,T_4^{(a)}\}$ and $\mathcal{B}$ is not a maximal or second weight configuration then, $$N(\mathcal{B})<N'_3.$$  \end{proposition}

\begin{preuve}Let $k$ and $d_1,\ldots,d_k$ defining an hyperplane arrangement $\mathcal{B}\in\mathcal{L}_d\setminus\{T_{max},T_4,T_1,T_2,T_4^{(a)}\}$ and $d'=\displaystyle\sum_{i=1}^kd_i=t'(q-1)+s'$, $0\leq s'\leq q-2$.
\begin{itemize}\item Assume there exists two distinct $1\leq i_1\leq k$ and $1\leq i_2\leq k$ such that $$1\leq d_{i_1}\leq d_{i_2}\leq q-2.$$
Then let replace an hyperplane of the block $i_1$ in $\mathcal{B}$ by a new hyperplane of the block $i_2$ in $\mathcal{B}$. We denote by $\mathcal{B}'$ this new hyperplane arrangement. Then, $N(\mathcal{B}')>N(\mathcal{B})$.
Since $\mathcal{B}$ is not a configuration of type $T_1$, $T_2$, $\mathcal{B'}$ is not a maximal configuration. Since $\mathcal{B}$ is not a configuration of type $T_4^{(a)}$, $\mathcal{B}'$ is not a $T_4$ configuration. So, in this case, $N(\mathcal{B})<N_3'$.
\item Assume there exists $1\leq i_1\leq k$ such that all $d_i$ but $d_{i_1}$ are 0 or $q-1$. Then since $\mathcal{B}$ is not a maximal configuration, either $t'<t$ or $t'=t$ and $s'=0$. So, we can add 1 hyperplane in a new direction. We denote by $\mathcal{B}'$ this new configuration and we have 
$N(\mathcal{B}')>N(\mathcal{B})$. By definition, $\mathcal{B}'$ is neither a maximal configuration nor a $T_4$ configuration. So, in this case, $N(\mathcal{B})<N_3'$.
\item Assume for all $1\leq i\leq k$, $d_i$ is either 0 or $q-1$. Then, since $t'\leq t\leq m-1$, we can add a new hyperplane in a new direction. We denote by $\mathcal{B}'$ this new configuration and  we have $N(\mathcal{B}')>N(\mathcal{B})$. Since $\mathcal{B}$ is not a $T_4$ configuration, $\mathcal{B}'$ is not a maximal configuration. By definition, $\mathcal{B}'$ is not a $T_4$ configuration. So, in this case, $N(\mathcal{B})<N_3'$.
\end{itemize}\end{preuve}
\bibliographystyle{plain}
\bibliography{C:/Users/Elodie/Dropbox/bibliothese}

\end{document}